\documentclass[sigconf, screen, nonacm]{acmart}

\usepackage{balance}
\usepackage{tabularx} 
\usepackage{booktabs}

\newcommand\vldbdoi{XX.XX/XXX.XX}

\newcommand\vldbavailabilityurl{http://vldb.org/pvldb/format_vol14.html}

\newcommand{\sys}{\mbox{\textsc{EKO}}\xspace}

\newcommand{\sampler}{\mbox{\textsc{Sampler}}\xspace}

\newcommand{\preprocessor}{\mbox{\textsc{Preprocessor}}\xspace}
\newcommand{\featureextractor}{\mbox{\textsc{Feature Extractor}}\xspace}

\newcommand{\encoder}{\mbox{\textsc{Encoder}}\xspace}
\newcommand{\decoder}{\mbox{\textsc{Decoder}}\xspace}

\newcommand{\dataloader}{\mbox{\textsc{Data Loader}}\xspace}
\newcommand{\filter}{\mbox{\textsc{Filter}}\xspace}
\newcommand{\udf}{\mbox{\textsc{User Defined Function}}\xspace}

\newcommand{\execution}{\mbox{\textsc{Execution Engine}}\xspace}

\newcommand{\tasti}{\mbox{\textsc{Tasti}}\xspace}
\newcommand{\noscope}{\mbox{\textsc{NoScope}}\xspace}
\newcommand{\uniform}{\mbox{\textsc{Uniform}}\xspace}
\newcommand{\iframe}{\mbox{\textsc{I-Frame}}\xspace}
\newcommand{\none}{\mbox{\textsc{No-Sampling}}\xspace}
\newcommand{\sysvgg}{\mbox{\textsc{EKO-VGG}}\xspace}

\newcommand{\tight}{\mbox{\textsc{Tight}}\xspace}
\newcommand{\loose}{\mbox{\textsc{Loose}}\xspace}
\newcommand{\medium}{\mbox{\textsc{Medium}}\xspace}

\newcommand{\midd}{\mbox{\textsc{Middle}}\xspace}
\newcommand{\mean}{\mbox{\textsc{Mean}}\xspace}
\newcommand{\first}{\mbox{\textsc{First}}\xspace}

\newcommand{\paperTitle}{EKO: Adaptive Sampling of Compressed Video Data} 
\newcommand{\paperAuthors}{Jaeho Bang,Joy Arulraj}
\newcommand{\paperKeywords}{Video Analytics; Video Compression; Sampling; Database Applications}

\newcommand{\tahoma}{\mbox{\textsc{TAHOMA}}\xspace}
\newcommand{\probabilistic}{\mbox{\textsc{Probabilistic Predicates}}\xspace}
\newcommand{\focus}{\mbox{\textsc{FOCUS}}\xspace}
\newcommand{\blazeit}{\mbox{\textsc{BLAZEIT}}\xspace}
\newcommand{\miris}{\mbox{\textsc{MIRIS}}\xspace}
\newcommand{\exsample}{\mbox{\textsc{EXSAMPLE}}\xspace}

\newcommand{\vstore}{\mbox{\textsc{VSTORE}}\xspace}
\newcommand{\lightdb}{\mbox{\textsc{LIGHTDB}}\xspace}
\newcommand{\tasm}{\mbox{\textsc{TASM}}\xspace}

\newcommand{\scarone}{\mbox{\textsc{Q1}}\xspace}
\newcommand{\scartwo}{\mbox{\textsc{Q2}}\xspace}
\newcommand{\ucartwo}{\mbox{\textsc{Q3}}\xspace}
\newcommand{\ucarthree}{\mbox{\textsc{Q4}}\xspace}
\newcommand{\uvanone}{\mbox{\textsc{Q5}}\xspace}

\setcopyright{none}
\acmDOI{}
\acmISBN{}
\acmYear{}
\copyrightyear{} 
\acmPrice{}
\acmConference[]{}{}{}

\definecolor{linkcolor}{HTML}{647382}
\definecolor{citecolor}{HTML}{647382} %
\definecolor{urlcolor}{rgb}{0.4,0.2,0.2}
\definecolor{sqlcolor}{HTML}{965d67}
\definecolor{smtcolor}{HTML}{5d968c}
\definecolor{commentcolor}{HTML}{4B0082}

\hypersetup{
    pdfauthor = {\paperAuthors},
    pdftitle = {\paperTitle},
    pdfkeywords = {\paperKeywords},
    bookmarksopen = {true},
    colorlinks=true,
    citecolor={urlcolor},
    linkcolor={linkcolor},
 	urlcolor={citecolor},
    pdfborder={ 0 0 0 }
}

\usepackage{amsmath,amsopn}
\usepackage{listings}
\usepackage{fancyvrb}
\usepackage{supertabular,booktabs}
\usepackage{array,underscore,relsize}
\usepackage{enumitem}
\usepackage{balance}
\usepackage{booktabs}
\usepackage{pifont}
\usepackage{multirow}
\usepackage[scaled]{beramono}
\usepackage{tabularx}
\usepackage{graphics}
\usepackage{subfig}

\usepackage[compact]{titlesec}

\newcounter{example}[part]

\usepackage{aliascnt}
\usepackage[ruled,linesnumbered,noend]{algorithm2e}

\SetCommentSty{mycommfont}

\usepackage{enumitem}

\SetAlFnt{\footnotesize}
\SetAlCapFnt{\small}
\SetAlCapNameFnt{\small}

\usepackage[capitalize,noabbrev,nameinlink]{cleveref}

\crefname{lstlisting}{listing}{listings}
\Crefname{lstlisting}{Listing}{Listings}

\usepackage[compact]{titlesec}
\titlespacing{\section}{0pt}{5pt}{2pt}
\titlespacing{\subsection}{0pt}{3pt}{1pt}
\titlespacing{\subsubsection}{0pt}{3pt}{1pt}

\usepackage{soul}

\usepackage[title]{appendix}

\newcommand{\cc}[1]{\mbox{\smaller[0.5]\texttt{#1}}}

\fvset{fontsize=\scriptsize,xleftmargin=8pt,numbers=left,numbersep=5pt}

\def\Snospace~{\S{}}

\newif\ifdraft\drafttrue
\newif\ifnotes\notestrue
\ifdraft\else\notesfalse\fi

\newcolumntype{R}[1]{>{\raggedleft\let\newline\\\arraybackslash\hspace{0pt}}p{#1}}

\newcommand{\PP}[1]{
\vspace{0.032in}
\noindent{\bf\textsc{#1}.}\xspace
}

\newcommand{\squishitemize}{
 \begin{list}{$\bullet$}
  { \setlength{\itemsep}{0pt}
     \setlength{\parsep}{3pt}
     \setlength{\topsep}{3pt}
     \setlength{\partopsep}{0pt}
     \setlength{\leftmargin}{1.95em}
     \setlength{\labelwidth}{1.5em}
     \setlength{\labelsep}{0.5em} } }

\newcounter{Lcount}
\newcommand{\squishlist}{
    \begin{list}{\arabic{Lcount}. }
   { \usecounter{Lcount}
        \setlength{\itemsep}{0pt}
        \setlength{\parsep}{3pt}
        \setlength{\topsep}{3pt}
        \setlength{\partopsep}{0pt}
        \setlength{\leftmargin}{2em}
        \setlength{\labelwidth}{1.5em}
        \setlength{\labelsep}{0.5em} } }

\newcommand{\squishend}{\end{list}}

\usepackage{xstring}

\newcommand{\eg}{\textit{e.g.,}\xspace}
\newcommand{\ie}{\textit{i.e.,}\xspace}

\makeatletter
\newcommand\BeraMonottfamily{%
  \def\fvm@Scale{0.85}
  \fontfamily{fvm}\selectfont
}
\makeatother

\definecolor{mymauve}{rgb}{0.58,0,0.82}

\lstdefinestyle{SQLStyle}{
  language=SQL,
  basicstyle={\small\ttfamily},
  breaklines=true,
  frame=none,
  numbers=none,
  keepspaces=true,
  captionpos=b,
  stringstyle=\color{mymauve},
  keywordstyle=\color{blue},
  commentstyle=\color{dkgreen},
}

\lstdefinestyle{ScriStyle}{
language=SQL,
basicstyle=\BeraMonottfamily\footnotesize, 
keywordstyle=\color{smtcolor}\bfseries,
commentstyle=\color{commentcolor}\ttfamily,
morekeywords={and, or, not},
literate = {-}{-}1, 
}

\lstdefinestyle{PythonStyle}{
language=Python,
basicstyle=\BeraMonottfamily\footnotesize, 
keywordstyle=\color{smtcolor}\bfseries,
commentstyle=\color{commentcolor}\ttfamily,
morekeywords={and, or, not},
literate = {-}{-}1, 
columns=fullflexible,
breaklines=true,
postbreak=\mbox{\textcolor{blue}{$\hookrightarrow$}\space},
}

\definecolor{eclipseStrings}{RGB}{42,0.0,255}
\definecolor{eclipseKeywords}{RGB}{127,0,85}
\colorlet{numb}{magenta!60!black}

\lstdefinelanguage{JSON}{
    basicstyle=\normalfont\ttfamily,
    commentstyle=\color{eclipseStrings}, 
    stringstyle=\color{eclipseKeywords}, 
    showstringspaces=false,
    breaklines=true,
    string=[s]{"}{"},
    comment=[l]{:\ "},
    morecomment=[l]{:"},
    literate=
        *{0}{{{\color{numb}0}}}{1}
         {1}{{{\color{numb}1}}}{1}
         {2}{{{\color{numb}2}}}{1}
         {3}{{{\color{numb}3}}}{1}
         {4}{{{\color{numb}4}}}{1}
         {5}{{{\color{numb}5}}}{1}
         {6}{{{\color{numb}6}}}{1}
         {7}{{{\color{numb}7}}}{1}
         {8}{{{\color{numb}8}}}{1}
         {9}{{{\color{numb}9}}}{1}
}

\lstdefinestyle{JSONStyle}{
language=JSON,
basicstyle=\BeraMonottfamily\footnotesize, 
keywordstyle=\color{smtcolor}\bfseries,
commentstyle=\color{commentcolor}\ttfamily,
morekeywords={and, or, not},
literate = {-}{-}1, 
columns=fullflexible,
breaklines=true,
postbreak=\mbox{\textcolor{blue}{$\hookrightarrow$}\space},
}

\crefname{lstlisting}{listing}{listings}
\Crefname{lstlisting}{Listing}{Listings}

\definecolor{webgreen}{rgb}{0,.5,0}                                                                                                                     
\definecolor{webbrown}{rgb}{.6,0,0}                                                                                                                   
\definecolor{webblue}{rgb}{0,0,.7} 

\usepackage{siunitx}                                                                                                                                  
\usepackage{tikz}

\begin{document}
  
\title{\paperTitle}


\author{Jaeho Bang}
\email{jaehobang@gatech.edu}
\affiliation{%
  \institution{Georgia Institute of Technology}
}
\author{Pramod Chunduri}
\email{pramodc@gatech.edu}
\affiliation{%
  \institution{Georgia Institute of Technology}
}
\author{Joy Arulraj}
\email{arulraj@gatech.edu}
\affiliation{%
  \institution{Georgia Institute of Technology}
}

\renewcommand{\shortauthors}{Bang, et al.}

\begin{abstract}
Researchers have presented systems for efficiently analysing video data at 
scale using sampling algorithms.
While these systems effectively leverage the temporal redundancy present in
videos, they suffer from three limitations.
First, they use traditional video storage formats are tailored for human
consumption.
Second, they load and decode the entire compressed video in memory
before applying the sampling algorithm.
Third, the sampling algorithms often require labeled training data obtained
using a specific deep learning model.
These limitations lead to lower accuracy, higher query execution time, and
larger memory footprint.

In this paper, we present \sys, a storage engine for efficiently managing video
data. 
\sys relies on two optimizations. 
First, it uses a novel unsupervised, adaptive sampling algorithm for 
identifying the key frames in a given video.
Second, it stores the identified key frames in a compressed representation 
that is optimized for machine consumption. 
%
%
We show that \sys improves F1-score by up to 9\% compared to the next best performing
state-of-the-art unsupervised, sampling algorithms by selecting more
representative frames.
It reduces query execution time by 3$\times$ and memory footprint by
10$\times$ in comparison to a widely-used, traditional video storage format.
\end{abstract}

\maketitle



\section{Introduction}
\label{sec:introduction}

The advent of high-definition cameras has led to an explosion in the amount of
generated video data~\cite{youtube2020,youtubehour}.
%
%
Researchers have proposed several techniques for efficiently analysing video
data at scale.
These efforts have focused on: 
(1) reducing query execution time,
(2) increasing query accuracy, and
(3) reducing memory footprint.
To accelerate query processing, researchers have proposed techniques for 
sampling and filtering data using lightweight
models~\cite{kang2017noscope,kang2018blazeit,lu2018pp,suprem2020odin}.    
Another line of research has focused on techniques for reducing the memory
footprint of a video database management system (VDBMS) using versioning and
content-based compression~\cite{haynes2018lightdb,daum2020tasm}.


\PP{Challenges}
Traditional video storage formats seek to store videos in a representation that
is better suited for human consumption.
For instance, videos are often stored using a high frame rate and compressed
using a delta encoding scheme that supports random access~\cite{wiegand2003h264}. 
These design decisions are based on the assumption that visual data is mainly
consumed by humans.
However, this representation is \textit{not} efficient when videos are
primarily analysed by machines.
With a high frame rate, there is significant redundancy of information across
nearby frames.
As shown in~\cref{fig:contiguousframes}, the contents of the video only evolve
slightly across a span of 4 contiguous frames.
This increases the memory footprint of the VDBMS and the time that it takes to
decode the video~\cite{kang2020jointly}.
For instance, a 30-minute video stored at a frame rate of 60~fps occupies 24~GB of
memory after it is loaded and decoded into a series of frames in memory by the
VDMS.
Optimizing the compressed video representation for machine consumption would
enable faster query processing and lower the memory footprint of the VDBMS.

Another challenge lies in the usability and efficacy of the sampling
algorithms in the VDBMS.
These algorithms suffer from two limitations.
First, they often require labeled training data obtained using a specific deep
learning model~\cite{kang2018blazeit,kang2020tasti,moll2020exsample}.
So, the sampling algorithm is constrained to a specific dataset.
Second, they are tailored for retrieving a static number of samples from a given
video.
So, the algorithm needs to tuned again for fetching a different number of
samples.
Designing a sampling algorithm to circumvent these limitations would improve the
accuracy of the VDBMS.




\begin{figure*}[t]
    \includegraphics[width=0.8\textwidth]{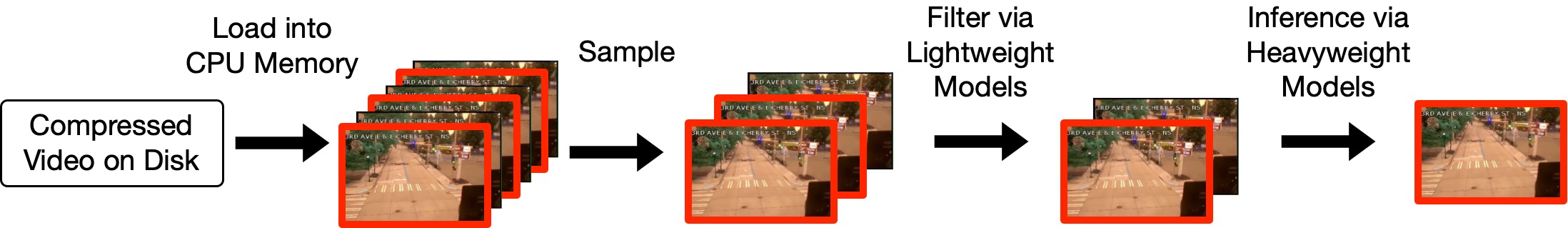}
    \newline
    \caption{
    \textbf{Video Analytics Pipeline} -- 
    The key components of a pipeline for processing a query over videos.
    }
    \label{fig:video-analytics-pipeline}
\end{figure*}

\PP{Our Approach}
In this paper, we present \sys, a storage engine for efficiently managing video
data.
\sys relies on two key techniques.
First, it uses a novel sampling algorithm for identifying the key frames in a
given video.
Second, it stores the identified key frames in a compressed representation that
is optimized for machine consumption with support for dynamic sampling rates.
Using these two techniques, \sys improves accuracy, reduces the query
execution time, and lowers the memory footprint.
%
%

\sys consists of three components: (1) \preprocessor, (2) \encoder, and (3)
\decoder (illustrated in~\cref{fig:systemarchitecture}).
First, the \preprocessor identifies the key frames in the given video
using a clustering algorithm.
It constructs a hierarchy of these representative frames to support dynamic
sampling rates.
Second, the \encoder stores the key frames in a compressed representation
tailored for machine consumption.
Third, the \decoder efficiently retrieves a subset of frames from the compressed
video based on the requirements of the \execution.

Our evaluation of \sys shows that it performs up to 9\% better in terms of F1-score than the next best
state-of-the-art unsupervised, sampling algorithms when selecting 0.1\% of the video for evaluation.
The gap between \sys and other methods significantly increases as we lower the number of frames sampled.
It reduces query execution time by 3$\times$ and memory footprint by 10$\times$
in comparison to a widely-used, traditional video storage format.
\sys generates compressed videos that increase the storage footprint compared to traditional compressed videos by 2$\times$.
But, this is still 100$\times$ smaller than that associated with storing the
video as a series of frames on disk to facilitate faster query processing.

\PP{Contributions}
The key contributions of this paper are:

\squishitemize
    \item We motivate the need for a compressed representation tailored for
    machine consumption (\autoref{sec:motivation}).
    \item We present a clustering algorithm for identifying the key
    frames of a video in an unsupervised manner (\autoref{sec:preprocess}).
    \item We introduce an adaptive sampling algorithm for efficiently decoding
    videos while processing queries (\autoref{sec:encoding-decoding}). 
    \item We demonstrate that \sys reduces query execution time, improves
    accuracy compared to other unsupervised sampling algorithms, and
    significantly reduces memory footprint in comparison to a traditional
    storage format (\autoref{sec:experiments}).
\squishend


\section{Motivation \& Background}
\label{sec:motivation}


We now present a motivating example that highlights the need for a
machine-centric compressed representation in~\autoref{sec:motivation::example}. 
We then provide a brief overview of related work on compressing, 
analysing, and sampling videos in~\autoref{sec:motivation::background}.

\subsection{Motivating Example}
\label{sec:motivation::example}
Consider the following query that retrieves frames with more than two cars from
a traffic surveillance dataset:

\begin{lstlisting}[style=SQLStyle]
SELECT frames 
FROM UA_DETRAC 
WHERE vehicle_type = 'car' 
  AND target_vehicle_count >= 2;
\end{lstlisting}

\begin{table*}[t!]
\centering
\small

\begin{tabular}
{c | c c c c c }
\toprule
\shortstack{\textbf{VDBMS}} & 
\shortstack{\textbf{Custom Feature} \\ \textbf{Extraction Network}} & 
\shortstack{\textbf{Sampling} \\ \textbf{Algorithm}} &
\shortstack{\textbf{Label Propagation} \\ \textbf{Methodology}} &
\shortstack{\textbf{Distance} \\ \textbf{Metric}} \\
\midrule

\sys & 
\checkmark &
Clusterer + Frame Selector &
Clusterer &
L2 &

\\

NOSCOPE &
        &
Thresholding        &
        &
L2      &

\\
TASTI &
\checkmark &
FPF & 
KNN &
L2 &
\\

US &
        &
Uniform Sampling &
        &
        &

\\ 
\bottomrule

\end{tabular}

\caption{ 
\textbf{Video Sampling Algorithms --}
Qualitative comparison of sampling algorithms used in recently proposed VDBMSs.
}
\label{tbl:sample-methods-overview}
\end{table*}

As shown in~\autoref{fig:video-analytics-pipeline}, a VDBMS answers this query
using the following operators. 
First, it loads all the frames of the given video from disk to CPU memory and
decodes them.
Second, it samples a subset of frames from the decoded, in-memory sequence
of frames.
The sampling algorithm takes advantage of the fact that the contents of nearby
frames in a video are similar to each other.
Third, it filters out a subset of the sampled frames using a lightweight 
machine learning model (\eg linear SVM \cite{cortes1995svm}).
While the filtering algorithm is less accurate than a deep learning model, it
lowers the query execution time by discarding a significant subset of irrelevant
frames.
Lastly, it moves the filtered frames from CPU memory and GPU memory and applies
the heavyweight deep learning model on these frames to evaluate the predicates
in the query (\eg type of vehicles present in the frame).

It is critical to answer the queries efficiently while satisfying the accuracy
requirements of the user.
A na\"ive, uniformly-random sampling algorithm hurts query accuracy by up to
100\% (\autoref{ssec:query-accuracy}).
In contrast, if the system were to sample infrequently for higher efficiency 
and the target event occurs rarely in the video, then it is likely that the
event goes undetected.
\sys tackles these challenges by identifying the key frames in the given video
using an unsupervised algorithm.
It then compresses the video in a machine-centric representation that keeps
track of the key frames.
Lastly, it decodes only the frames necessary for query processing to lower the
I/O overhead. 

\begin{figure*}[t]
    \includegraphics[width=0.9\textwidth]{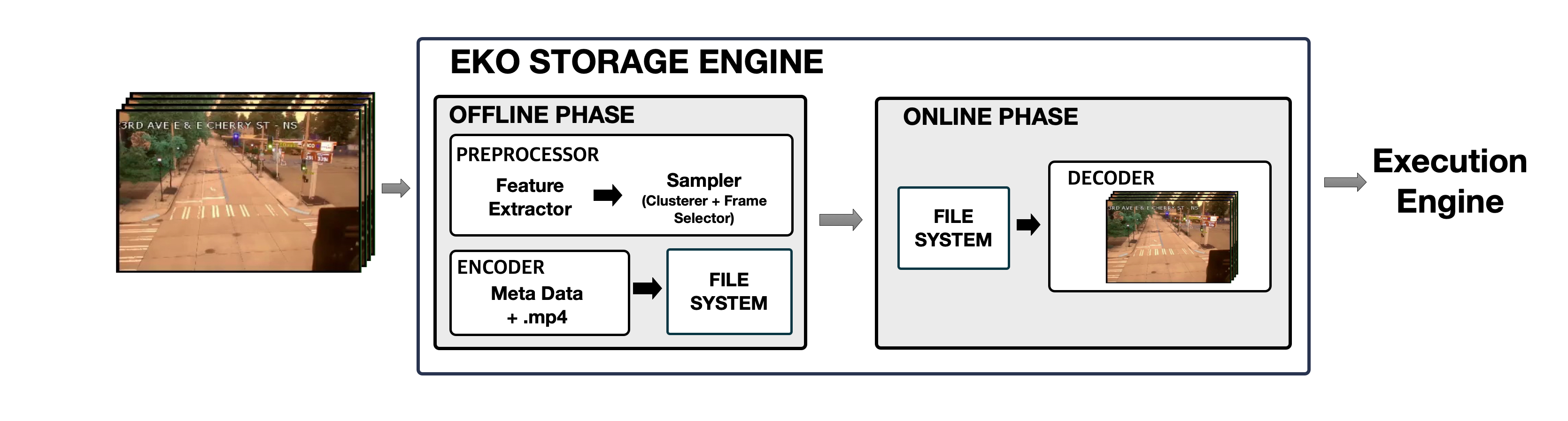}
    \newline
    \caption{
    \textbf{Architecture of \sys} -- 
    \sys consists of three components.
    The preprocessor extracts the key frames from the given video using an
    unsupervised sampling algorithm.
    The encoder utilizes these key frames to generate a differently compressed
    video.
    The decoder allows \sys to quickly load a subset of the key frames from the
    compressed video on disk to CPU memory for query execution.
    }
    \label{fig:systemarchitecture}
\end{figure*}

\subsection{Background}
\label{sec:motivation::background}

\begin{figure}[t]
    \includegraphics[width=0.45\textwidth]{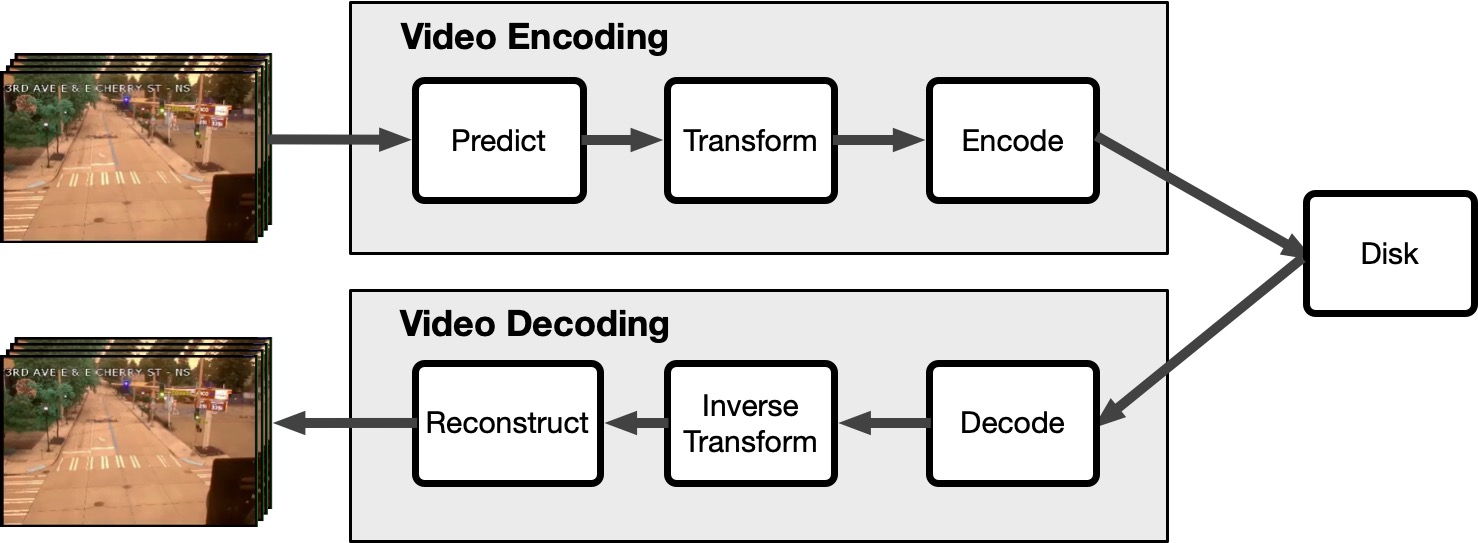}
    \caption{
    \textbf{H.264 Video Coding Format} --
    Steps associated with encoding and decoding H.264 video coding format.
    }
    \label{fig:videocompression}
  \end{figure}

\PP{Video compression}
Video compression is centered around: (1) intra-frame, and (2) inter-frame coding.
Intra-frame coding leverages correlation between pixels within the same frame.
Intra-frame coding process is identical to standard image compression standards such as JPEG compression.
As shown in~\autoref{fig:videocompression}, it consists of the following steps: 
(1) prediction, 
(2) transformation, 
(3) quantization and encoding.
Frames are initially divided into \textit{macroblocks} (typically 16x16 pixels) 
and content predictions are made using surrounding \textit{macroblocks}.
\textit{Macroblock} values are transformed using discrete cosine transform functions (DCT).
The coefficients of the DCT are quantized according to the specified precision.
Intra-frame coding compresses video by $3\times$ compared to raw pixel values
(\autoref{ssec:storage}) and enables fast random access to a particular frame.
Inter-frame coding leverages the temporal redundancy between nearby frames in
the video.
It uses the same macroblock used for intra-frame coding.
But, during the prediction step, it also uses information in previously
encoded frames.
The three major \textit{frame types} in a compressed video are:
(1) I-frames that serve as reference frames and do not require other frames for
decoding, 
(2) P-frames require data from previous frames for decoding, and 
(3) B-frames require data from both previous and forward frames for decoding.
We refer to P- and B-frames as \textit{delta frames} since they only store the
difference between nearby frames rather than each full frame.
\cref{fig:ibp-frames} illustrates the differences between I and P frames.
We elaborate on how \sys encodes a different set of I-frames
in~\autoref{sec:encoding-decoding::eko}. 

The \encoder first determines how to best represent the current frame of
interest: I- or B- or P-frame.
The \encoder generates motion vectors to capture the movement in video with
respect to prior frames.
It calculates the difference between the prediction for the current frame and
its actual contents, and encodes the difference as a residual.
Inter-frame coding further decreases the storage footprint of the video by
$100\times$ (\autoref{ssec:storage}).

Video container formats often use both intra- and inter- frame coding 
(\eg AVI~\cite{avi1998}, MP4 ~\cite{pereira2002mpeg}).
The downside to this approach is that random access takes time.
This is because when a VDBMS seeks to extract a certain frame in the video, 
the decoder must first access the first frame in the group of pictures (GOP) 
and apply the delta calculations on top of that frame.
\sys manipulates the I-frames of the video during ingestion to accelerate query
processing and to lower the storage footprint.

\PP{Video analytics pipeline}
As shown in~\autoref{fig:video-analytics-pipeline},  in contemporary VDBMS
systems, the query processing pipeline consists of the following components:
(1) \dataloader, (2) \sampler, (3) \filter, and an
(4) \udf (UDF).
The \dataloader migrates the video data from disk to CPU memory.
It only loads the segments of the video that are required for processing the
query.
The \sampler is responsible for selecting a set of important frames in the loaded
video.
A couple of na\"ive sampling algorithms are:
(1) uniformly sampling frames from the video, and
(2) selecting the I-frames in the compressed video.
Researchers have proposed sampling algorithms that better leverage the temporal
redundancy of the video~\cite{moll2020exsample,jiang2018chameleon,kang2017noscope}.

UDF refers to the heavyweight deep learning model used by the VDBMS to process
the query.
This includes object detection (\eg YOLO \cite{redmon2016yolo}) and
object classification models (\eg VGG \cite{simonyan2014vgg}).
\filter are faster models compared to UDFs.
However, their performance comes at the expense of a drop in accuracy.
The \filter take a variety of forms ranging from traditional machine learning
models (\eg linear SVM, random forest) to specialized deep neural networks.
VDBMSs leverages filters to accelerate queries with a tolerable accuracy
drop~\cite{lu2018pp, anderson2019tahoma}.

\begin{figure*}[t]
    \includegraphics[width=0.85\textwidth]{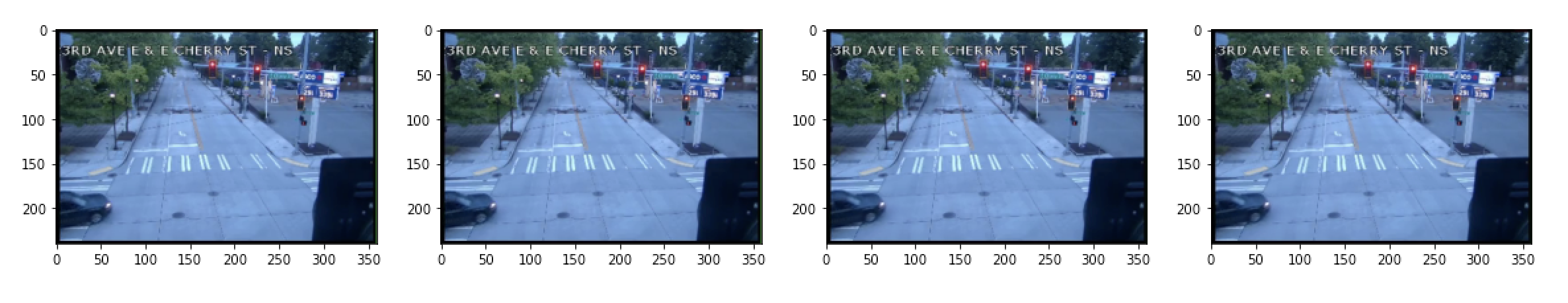}
    \caption{
    \textbf{Temporal Redundancy} -- 
    Corresponding pixels in contiguous video frames tend to take similar values.
    Video compression schemes exploit this temporal redundancy by only storing
    the differences between nearby frames rather than each full frame.
    }
    \label{fig:contiguousframes}
\end{figure*}

\PP{Video Sampling}
VDBMSs apply sampling algorithms to leverage the temporal redundancy in videos.
Depending on the sampling rate, the VDBMS may answer the query orders of
magnitudes faster.
\blazeit~\cite{kang2018blazeit} utilizes control variates to efficiently sample
video frames for \cc{LIMIT} queries.
\exsample~\cite{moll2020exsample} is tailored for queries that require detection
of distinct objects.
Based on the distribution of objects, it adopts a different sampling technique.
These algorithms require labeled data for sampling frames.

Researchers have also presented unsupervised, sampling algorithms.
\noscope~\cite{kang2017noscope}, \tasti~\cite{kang2020tasti}, and \sys fall
under this category.
We present a qualitative comparison of these algorithms
in~\cref{tbl:sample-methods-overview}.
\noscope samples frames whenever the difference between
the current frame and the previous frame exceeds a certain threshold.
Since there is no specified method of propagating labels to the frames that have
not been sampled, the set of queries that it can answer is limited.
It does not rely on a feature extraction network to derive the features for a
given image.
\tasti is the closest system in terms of objectives to \sys.
It leverages a custom feature extraction network based on triplet loss~\cite{chechik2010triplet}. 
It selects samples using the Furthest Point First (FPF)
algorithm~\cite{rosenkrantz1977fpf}, and 
propagates labels using a variation of K-Nearest Neighbors (KNN)
algorithm~\cite{altman1992knn}. 
We defer a comparison of \sys against these systems
to~\autoref{ssec:query-accuracy}.


\section{System Overview}
{\label{sec:systemarchitecture}}

The architecture of \sys is shown in~\cref{fig:systemarchitecture}. 
\sys seeks to:
(1) accelerate inference queries, and 
(2) minimize the CPU memory footprint during query processing.
It automatically extracts the key frames in the given video using an
unsupervised algorithm.
The benefits of this design decision are twofold.
First, most of the video content being created do not have labels associated
with them.
\sys supports such unlabeled data.
Second, it exploits the innate structure of videos thereby allowing it to
support a broader range of UDFs and queries.
%

\textbf{Use Case}. 
Consider a database with a week's worth of video data collected from a camera
deployed at a given intersection.
If an urban planner is counting the number of cars passing by this intersection,
\sys quickly answers this query by sampling 1\% of the frames and
processing them using the rest of video analytics pipeline.
To facilitate efficient sampling, \sys pre-processes the video data in an 
\textit{offline} manner to pick the most representative frames during ingestion
and stores the processed videos on disk.
When the user issues the query, it only decodes the top 1\% of the frames that
summarize the given dataset, and sends it to the \execution.

\textbf{Workflow.} 
When the user loads a video into the VDBMS, \sys summarizes the given data.
\sys's \preprocessor automatically infers the optimal number of samples using
the Silhouette technique~\cite{rousseeuw1987silhouettes}.
The \encoder stores the sampling results in the compressed representation.
Lastly, when the user issues a query, the \decoder loads the necessary
frames from the compressed video to answer the query efficiently.


\begin{algorithm}[t]
\SetAlgoLined
\DontPrintSemicolon
\SetKwFunction{FPreprocess}{Preprocess}
\SetKwFunction{FExtractFeatures}{ExtractFeatures}
\SetKwFunction{FClusterFeatures}{ClusterFeatures}
\SetKwFunction{FSelectFrames}{SelectFrames}
\SetKwProg{Fn}{Function}{:}{}
\SetKwInOut{Input}{Input}
\SetKwInOut{Output}{Output}
\Input{Decompressed video dataset $M$ (\eg frames) \newline 
        Number of samples $N$ (\emph{optional})}
\Output{Set of sampled frames $S$ (subset of $M$)}

\Fn{\FPreprocess{$M$, $N$}}{
\label{alg:eko-pipeline::preprocess}
$D = ExtractFeatures(M)$ \\
$C = ClusterFeatures(D, N)$ \\
$S = SelectFrames(C)$ \\
 
\Return $S$
 }

\Fn{\FExtractFeatures{$M$}}{
  \label{alg:eko-pipeline::extractfeatures}
$FE = InitFeatureExtractionNetwork()$ \\
\For{$M_i \in M$}{
    \tcp{Downsample the video frames using EKOD}
	$D_i = FE(M_i)$ \;
 
 } 
\Return $D$
}

\Fn{\FClusterFeatures{$D$, $N$}}{
  \label{alg:eko-pipeline::clusterfeatures}
  \tcp{Compute connectivity matrix to enforce explicit temporal constraint}
  $conn = ComputeConnectivity(D)$\;
  \tcp{Compute optimal number of samples using heuristics if not given}
  \eIf{$N$ is $NONE$}{
    $N = ComputeOptimalN(D)$\;
    }{}
  
  \tcp{Cluster based on features and connectivity}
  $C = Cluster(D, conn, N)$\;

  \Return $C$
}

\Fn{\FSelectFrames{$C$}}{
  \label{alg:eko-pipeline::selectframes}
  \tcp{Sample a frame from each cluster}
  \For{$C_i \in C$}{
     $S_i = SelectTemporalMiddleFrame(C_i)$\;
  }
  \Return $S$
}

 \caption{Algorithm for sampling frames in a video dataset.}
 \label{alg:eko-pipeline}
\end{algorithm}

\section{Preprocessor}
{\label{sec:preprocess}}

The \preprocessor is based a novel, unsupervised clustering algorithm 
that is tailored for the given video dataset and that iteratively converges to 
an optimal set of cluster centers.
It finds the set of representative frames from the given video that is used for
subsequent query processing.
The benefits of the \preprocessor are twofold.
First, it outperforms other widely-used sampling algorithms 
(\eg uniform sampling), especially for queries focusing on rare events. 
Second, it does not require labeled video data.

\begin{figure}[t]
    \includegraphics[width=0.45\textwidth]{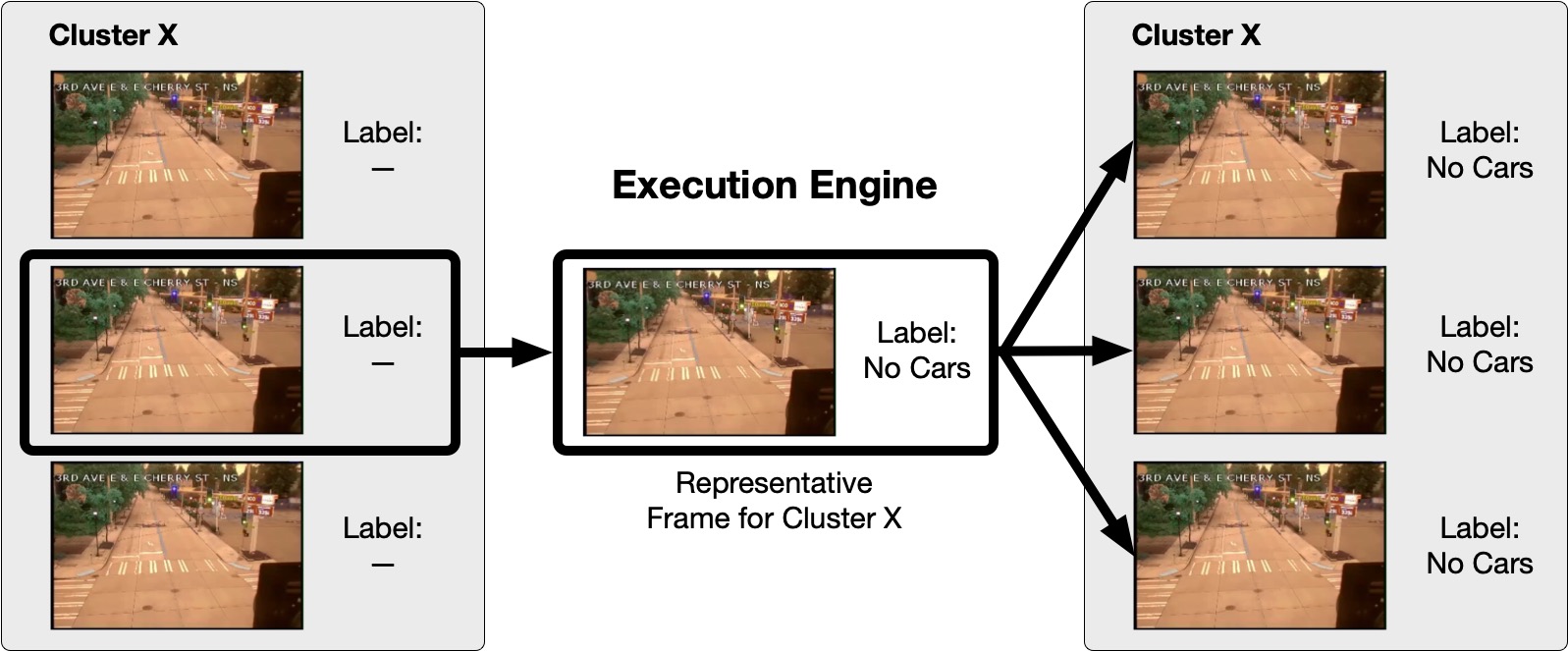}
    \newline
    \caption{\textbf{Label Propagation Method} -- 
    The \execution propagates the binary label assigned to a sampled key frame
    to the other frames within the same cluster.
    For instance, if the key frame contains no cars, it assumes that the other
    frames in the same cluster also contain no cars. 
    }
    \label{fig:label-propgation}
\end{figure}

As shown in~\cref{fig:label-propgation}, while processing the selected frames
using the \udf, the \execution propagates the binary label assigned to a
sampled frame to the other frames within the same cluster (that were not
selected by the sampling algorithm).
As illustrated in~\cref{fig:systemarchitecture}, the \preprocessor consists of
two components:
(1) \featureextractor, and 
(2) \sampler. 
We next describe these components in detail.

\subsection{Feature Extractor}

\cref{alg:eko-pipeline} outlines the steps for choosing the representative
frames.
The \preprocessor first downsamples the frames in the given video before
applying the clustering algorithm to circumvent the curse of
dimensionality~\cite{bellman1954theory}.
With high dimensional data, the L2 distance metric used for clustering the data
does not accurately capture the underlying distribution.
Lowering the dimensionality of the data increases the efficacy and the
efficiency of the clustering algorithm.
Since the complexity of the clustering algorithm grows quadratically with
the number of dimensions ($n$), reducing $n$ by 2$\times$ leads to a 4$\times$
speedup.

\PP{Challenges}
Choosing the appropriate downsampling algorithm is a non-trivial task.
As we later show in~\autoref{ssec:query-accuracy}, the downsampling
algorithm has a significant effect on the accuracy of the query.
%
The \featureextractor must satisfy the following objectives:
\squishitemize
    \item The \featureextractor must not only consider the raw pixel content of the
    frames, but also take the key \textit{objects} themselves into
    consideration.
    \item The \featureextractor must consider the temporal dynamics of the video.
    \item The downsampling algorithm must work well with the subsequent clustering
    algorithm.
\squishend

The reason why we need to focus on the objects is because the queries we target
are centered around them.
If the \featureextractor only considers the raw pixel content for clustering,  
it has a higher likelihood of not capturing the crucial features.

\PP{Our Approach}
\featureextractor uses a modified VGG16~\cite{simonyan2014vgg} network for feature extraction.
It implicitly enforces an ordering constraint to force similar frames in the
video to be grouped together after the clustering algorithm.
To embed this constraint, it fine-tunes the pretrained VGG16 network as
outlined in~\cref{alg:ekod-training}.
After extracting the features, it passes them to the clustering algorithm.
While training the model, it uses the labels generated from executing the 
clustering algorithm from the previous iteration to compute the loss.
\featureextractor uses the Adam optimizer for gradient-based optimization of the
network.

%
%
\sys{'s} training procedure is a variant of the canonical iterative parameter
update technique.
In particular, it is related to Deep Embedded Clustering (DEC)~\cite{dec}.
While DEC is tailored for images, \sys focuses on sampling from videos.
So, it uses a different clustering algorithm.
While DEC utilizes K-means for deriving the groups and to compute the loss, 
\sys relies on hierarchical clustering~\cite{ward1963hierarchical}.
Furthermore, \sys{'s} clustering algorithm enforces explicit temporal
constraints, thereby altering the behavior of the \featureextractor.

\begin{algorithm}[t]
\SetAlgoLined
\DontPrintSemicolon
\SetKwFunction{FTrain}{TrainFeatureExtractor}
\SetKwProg{Fn}{Function}{:}{}
\SetKwInOut{Input}{Input}
\SetKwInOut{Output}{Output}
\Input{Decompressed video dataset $M$ (\eg images) \\
       Number of samples $N$ (\emph{optional}) \\
       Iteration limit $I$ (\emph{default value: 100})}
\Output{Set of sampled frames $S$ (subset of $M$)}

\Fn{\FTrain{$M$, $N$}}{

$I_{curr}$ = 0\\
\While {$I_{curr} \le \mathcal{I}$}
{
 
 \hyperref[alg:eko-pipeline::extractfeatures]{$D = ExtractFeatures(M)$}\;
 \hyperref[alg:eko-pipeline::clusterfeatures]{$C = ClusterFeatures(D, N)$}\;
 \For{$D_i \in D$}{
    \tcp{For the current features, find the selected frame within the cluster}
    $S_j = FindClusterRep(D_i)$\;
    \tcp{Compute the distance between the selected frame and current frame}
    $L = ComputeDist(D_i, S_j)$\;
    \tcp{Perform backpropagation to update the network's parameters}
    $L.Backprop()$
 }
 }

\Return $\mathcal{S}$
}

\caption{Algorithm for training the feature extractor.}
\label{alg:ekod-training}
\end{algorithm}

\subsection{Sampler}
\label{sec:sampler}


As shown in~\cref{alg:eko-pipeline}, the \sampler takes the downsampled frames
returned by \featureextractor as input.
It first clusters these frames together and then selects a representative frame
from each cluster.
We modify the canonical hierarchical clustering
algorithm~\cite{ward1963hierarchical}.
First, \sys explicitly adds a \textit{temporal constraint} to group contiguous
frames together.  
Second, \sys automatically derive the required number of clusters by tailoring 
the silhouette method~\cite{rousseeuw1987silhouettes}.
Third, we empirically show that it is important to select the middle frame of
each temporal cluster as the representative frame.
\PP{Temporal Constraint}
While the \featureextractor chooses important features needed to understand the content
of the video frames, we found that it is important to explicitly add a 
temporal constrain to group contiguous frames together.
This improves the efficacy of the \sampler
(\autoref{ssec:temporal-constraint}).
Consider the movement of the car in~\cref{fig:contiguousframes}.
If a frame at timestamp $t$ contains one car, then it is likely that the frames
at timestamps $t-1$ and $t+1$ also contain one car.
By explicitly enforcing the clustering algorithm to consider groupings of
contiguous frames together, \sampler fully leverage the temporal redundancy 
in videos.

\PP{Dynamic Sample Selection}
\sys allows the user to specify the number of samples required for answering the
query.  
During ingestion, the \sampler computes the hierarchical tree specifying the
clusters and caches it.
Using this cached tree, it determines the cluster formations for the desired
number of samples.
Besides the already selected representative frames, it obtains additional
samples by selecting frames that are closest to the temporal median of each
cluster.
As we show in~\autoref{ssec:query-accuracy}, these samples deliver higher
accuracy compared to uniform sampling over the already selected representative
frames.  

\PP{Frame Selection}
After determining the optimal number of clusters and clustering the downsampled
datapoints, \sampler is left with the task of selecting the representative
frames of each cluster that will be then be sent to the \execution.
We take the temporal redundancy of videos into consideration in the frame
selection decision.
In particular, we assume that the probability that frames at timestamps $t$ and 
$t+1$ are likely to have the same label compared to that of frames separated by
more than 10 frames.
So, if \sampler must only pick one frame from each cluster, then selecting the
middle frame increases the likelihood that the other frames in the cluster
share the same label as the selected frame.
We show the importance of picking the middle frame
in~\autoref{ssec:frame-selection}.

\subsection{Theoretical Analysis}

We now present a theoretical analysis of the clustering algorithm used by \sys.
Consider a set of $n$ points in the set $\{x_i \in X\}^{n}_{i = 1}$ that we
seek to group together into $k$ clusters.
We assume that for every given dataset $X \in \mathbb{R}^{d_x}$, 
there exists an underlying latent space $Z \in \mathbb{R}^{d_z}$ that is 
optimal for sampling.
$X$ and $Z$ differ in that the latter dataset focuses more on the objects 
in the frames that are relevant for the video analytics queries.
While we cannot directly derive $Z$, we seek to learn 
$\hat{Z} \in {\mathbb{R}^{d_{\hat{z}}}}$ that approximates $Z$.
We denote the transformation function by $h_\theta : X \rightarrow \hat{Z}$,
where $\theta$ are learnable parameters. 

\PP{Characteristics of Latent Space}
We now list the desired characteristics of $\hat{Z}$.
First, $d_{\hat{z}} << d_x$.
Reducing the dimensionality of the dataset will allow \sys to circumvent the 
curse of dimensionality.
Next, if frames are temporally close to each other, they should be close to
each other in the latent space as well.
This property allows us to leverage the temporal redundancy of videos.
Lastly, every cluster in the latent space must be dense.
This ensures that frames with similar content are grouped together, thereby
increasing the likelihood of these frames being grouped together by the
clusterer.

\PP{Learning the Transformation Function}
To learn the transformation function $h_\theta$ in an unsupervised manner, 
we impose explicit and implicit constraints on the \featureextractor and its
training process. 
We begin with a pre-trained VGG16 network so that the \execution may leverage
the features returned by this network.
A deep learning network learns a function $f(x)$ of the form $f(x) =
g(h_\theta(x))$.
Here, $h_\theta(x)$ is the feature extraction network and $g(x)$ is the labeling
function.
We constrain $h_\theta$ to match the underlying distribution $\hat{Z}$ in two
ways.
First, we augment $\{\hat{z}_i \in {\hat{Z}}\}_{i=1}^n$ by concatenating 
the temporal location of the frame and downsizing the output of the VGG16
network using an fully connected layer.
This imposes an implicit temporal connectivity constraint on the points in the
latent space.
By downsizing the output of the VGG16 network, \sys chooses the key features,
and reduces the dimensionality of the dataset (\ie $d_{\hat{z}} << d_x$).
Second, we impose an additional explicit constraint for temporal connectivity.
To meet the desired characteristics of the latent space, we seek to find
$\theta$ that minimizes the following objective function:

$$ \min_{\theta} \Sigma_{i = 1}^n (f_\theta(x_i) - f_\theta(c(x_i))^2) $$

In other words, \sys tries to minimize the difference between the features of a
given video frame and that of the representative frame of the cluster to which
it belongs.

%
%
%


%
%


\section{Video Encoder and Decoder}
{\label{sec:encoding-decoding}}

\begin{figure}[t]
  \includegraphics[width=0.4\textwidth]{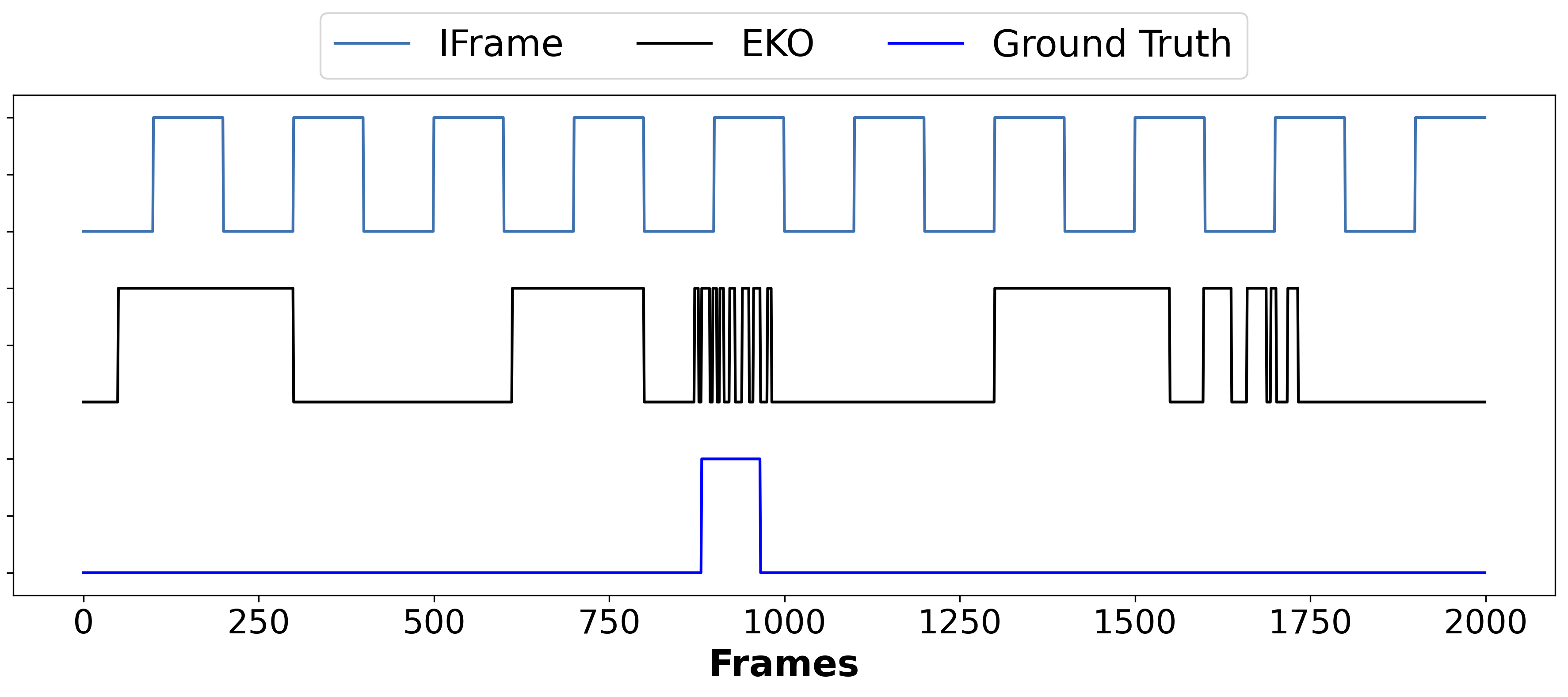}
  \caption{
  \textbf{Cluster Boundaries} --
  We represent each cluster by a transition along the sequence of frames.
  \sys adjusts the cluster boundaries based on the contents of the frames.
  }
  \label{fig:cluster-groupings}
\end{figure}

In this section, we first explain why canonical I-frames do not meet the
requirements of the VDBMSs in~\autoref{sec:encoding-decoding::eko}.
We then discuss how the frames sampled by \sys are different from 
I-frames in~\autoref{sec:encoding-decoding::analysis}. 
Lastly, we discuss how \sys encodes and decodes video in
~\autoref{sec:encoding-decoding::how}.

\subsection{Canonical I-Frames}
\label{sec:encoding-decoding::eko}

\begin{figure}[t]
  \includegraphics[width=0.4\textwidth]{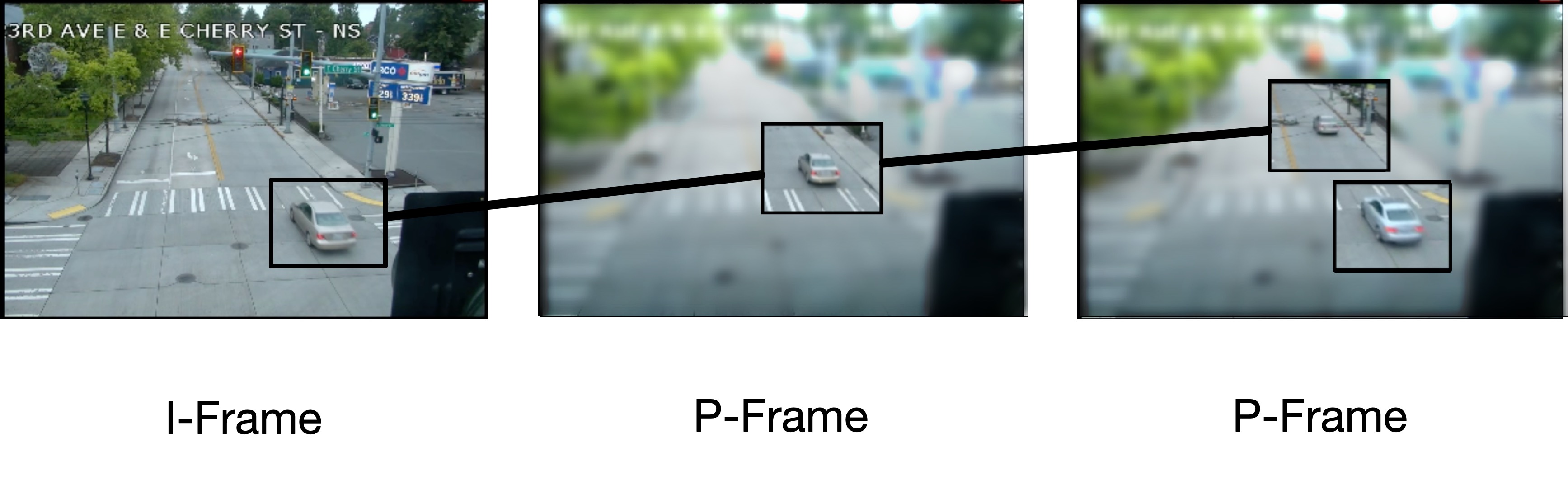}
  \caption{
  \textbf{I- and P-frames} -- 
  Visualization of intra-coded (I) and predicted (P) frames in a compressed
  video.
  }
  \label{fig:ibp-frames}
\end{figure}

In this paper, we focus on the H.264 video compression standard.
As shown in~\cref{fig:ibp-frames}, within a GOP, the I-frame is always the
first frame.
Subsequent P- and B-frames refer back to this I-frame for decoding
(\autoref{sec:motivation::background}).
The number of frames within a single GOP depends on multiple factors, including:
(1) frame rate, and (2) constant rate factor.
Typically, videos with higher frame rate have a GOP with around 250 frames.
Since I-frames do not refer back to any frame, it takes less time to decode a
random I-frame as opposed to other types of frames.
So, \sys seeks to use I-frames for processing the query.
In particular, the \execution propagates the labels returned by the object
detector on an I-frame to other nearby frames within the video.

\PP{Challenges}
The I-frames chosen by a canonical \encoder do not satisfy the requirements of
the \execution.
If \sys were to use them for processing the query, the drop in query accuracy is
significant (\autoref{ssec:query-accuracy}).
The reasons for this is twofold.
First, the canonical I-frames do not take the dynamics of the video into
consideration.
The size of the GOP is hard-coded to be a small constant (\eg 250 frames) 
to facilitate faster retrieval of other frames within the GOP.
Since I-frames typically occur at the beginning of each GOP, this is equivalent
to taking uniformly random samples from the video.
This approach does not work well for queries focusing on rare events.
Second, as we discussed in~\autoref{sec:preprocess}, the representative frame
should be the middle frame of the group (as opposed to the first frame).
This increases the likelihood that the other frames in the group share the
same label as the selected frame.
Given these constraints, the \encoder in \sys picks a different set of frames,
as described in~\autoref{sec:sampler}, to be the I-frames in the compressed
video.

\subsection{Sampled Frames}
\label{sec:encoding-decoding::analysis}

We now examine the differences between the traditional I-frames and the frames
sampled by \sys.
Since the \encoder uses these sampled frames as I-frames in the re-encoded
video, we examine the differences between these two set of frames.
We focus on the query shown in~\autoref{sec:motivation::example} for this
analysis.
We consider a dataset with 100~K frames and generate 1000 clusters.
We compare the groupings with traditional I-frames (\ie GOPs) and the clusters
generated in \sys.
We compute inter-cluster metrics to examine the relevance of frames
grouped together.
Unlike \sys, the locations of the I-frames are fixed and do not take the
activity in the video into consideration.
So, the clusters are uniformly sized.
In contrast, \sys generates clusters that exhibit a high variance in cluster
size.

\begin{table}[t]
    \begin{tabular}{ccc}
    \hline
                                & \multicolumn{2}{c}{\textbf{Inter Cluster Statistics}} \\ 
    \multicolumn{1}{c}{}       & Iframe                 & EKO                          \\
    \hline
    \multicolumn{1}{c|}{Mean}   & 100                    & 100                         \\
    \multicolumn{1}{c|}{Median} & 100                    & 26.5                     \\
    \multicolumn{1}{c|}{Std}    & 0                      & \textbf{120}                        \\
    \multicolumn{1}{c|}{Min}    & 100                    & \textbf{2}                        \\
    \multicolumn{1}{c|}{Max}    & 100                    & \textbf{706}                  \\   
    \hline
    \end{tabular}

    \caption{\textbf{Comparison of traditional I-frames and \sys --}
    Inter-cluster statistics associated with the clusters generated
    by a traditional encoder and \sys{'s} encoder. 
    }
    \label{tbl:cluster-statistics}
\end{table}

The results are shown in~\cref{tbl:cluster-statistics}.
With traditional I-frames, the standard deviation of the cluster size is 0.
In contrast, with \sys, the standard deviation of the cluster size is 120.
This is because \sys adjusts the cluster boundaries based on the variation in
the contents of the video.
The benefits of this technique is more prominent when the query is evaluated
using a fewer number of frames.
\cref{fig:cluster-groupings} illustrates the differences in the generated
cluster boundaries.
While groupings generated using traditional compression is static,  
\sys dynamically adjusts the clusters.
Due to correlation between changes in the contents of the video and changes in
the ground truth label, \sys delivers higher accuracy than traditional
I-frames.  

\subsection{Encoding and Decoding}
\label{sec:encoding-decoding::how}

\sys leverages the widely-used \textsc{ffmpeg} library~\cite{tomar2006ffmpeg}
for manipulating the video.
In particular, it uses the x264 codec to directly configure a custom set of
frames returned by \sampler as the I-frames in the re-encoded video.
It uses \textsc{ffmpeg} in this manner:

\begin{lstlisting}[language=SQL,
  showspaces=false,
  basicstyle=\ttfamily,
  keywordstyle=\color{blue},
  morekeywords={ffmpeg},
  commentstyle=\color{gray}]
ffmpeg -i src -force_key_frames T dst -y
\end{lstlisting}

Here, $T$ represents the indices of the sampled frames within the original
video.
$src$ and $dst$ represent the filenames of the original video and re-encoded
video, respectively.
While this re-encoding allows \sys to improve query accuracy, it also increases 
the storage footprint of the video by 80\% on average (\autoref{ssec:storage}).
%
%

%
\sys uses a custom \decoder that works in tandem with the \encoder.
Based on the query, it either decodes a subset of the I-frames or the entire
video.
We empirically show that the \decoder takes up to 10$\times$ less time to load 
a subset of frames picked by the \sampler in comparison to a canonical \decoder
operating on the original video (\autoref{ssec:query-speed}).


\section{Implementation}
\label{sec:implementation}

We implemented \sys in Python 3.
The \execution utilizes the Pytorch framework (v 3.7.6) for inference
using deep learning models.
The \encoder and \decoder leverage the ffmpeg framework (v 4.3) for
compressing and decompressing videos.
\sys analyzes videos in two stages.
\squishitemize
\item In the offline stage, the \preprocessor first loads the video using the 
OpenCV framework (v 4.2.0).
It then extracts key features from the video using the Pytorch framework.
Next, the \sampler takes the downsampled frames and applies a variant of the
agglomerative clustering algorithm in the scikit-learn library
(v 0.23.1).
In particular, it uses the \texttt{ward} linkage criterion to minimize the
variance of the clusters being merged.
The \sampler then selects a set of representative frames from these clusters.
These samples are passed on to the custom \encoder.
The \encoder re-encodes the video to embed the sampled frames as I-frames to
facilitate faster retrieval of these frames by the \execution during query
processing.

\item 
In the online stage, the \execution requests the \decoder to return a subset of
sampled frames.
The custom \decoder efficiently retrieves these frames from the re-encoded
video.
Lastly, the \execution applies inference on these representative frames.
It then propagates the labels returned by the deep learning on each
representative frame to other frames within the same cluster in the the
re-encoded video.
\squishend

\section{Experiments}
{\label{sec:experiments}}

In our evaluation, we illustrate that:
\squishitemize
    \item \sys delivers accurate results comparable to other sampling algorithms (\autoref{ssec:query-accuracy}).
    \item \sys's components contribute to the overall improvement in latency and accuracy (\autoref{ssec:ablation-study}).
    \item \sys executes queries up to 27$\times$ faster than traditional efficient sampling algorithms (\autoref{ssec:query-speed}).
    \item \sys lowers the storage footprint by 100$\times$ compared to a canonical storage format (\autoref{ssec:storage}).
    \item We conduct a sensitivity analysis of key parameters (\autoref{ssec:temporal-constraint}, \autoref{ssec:frame-selection}).
    
\squishend

\subsection{Evaluation Metrics}
\label{sec:baselines}

\PP{Latency} 
We measure the time it takes for \sys to load in the video from disk and to
process the query using the \execution~\cite{kang2020jointly}.
\PP{Accuracy}
To evaluate the accuracy of the sampling algorithms, we use a label propagation
technique.
In particular, \sys propagates the labels assigned by the \udf to the sampled
frame $X$ to other frames that are temporally close to $X$ in the video.
We compare these labels against the ground truth to measure precision and recall
metrics.

\subsection{Experimental Setup}


\begin{table}[]
    \begin{tabular}{ll}
    \toprule
    \textbf{Q1} & 
    {\begin{lstlisting}[style=SQLStyle]
    SELECT frames FROM Seattle 
    WHERE vehicle_type = 'car' 
      AND target_vehicle_count >= 1;
    \end{lstlisting} }  \\ 
    \textbf{Q2} & 
    {\begin{lstlisting}[style=SQLStyle]
    SELECT frames FROM Seattle 
    WHERE vehicle_type = 'car' 
      AND target_vehicle_count >= 2;
    \end{lstlisting}}   \\ 
    \textbf{Q3} & 
    {\begin{lstlisting}[style=SQLStyle]
    SELECT frames FROM UA-DETRAC 
    WHERE vehicle_type = 'car' 
      AND target_vehicle_count >= 2;
    \end{lstlisting}} \\ 
    \textbf{Q4} & 
    {\begin{lstlisting}[style=SQLStyle]
    SELECT frames FROM UA-DETRAC 
    WHERE vehicle_type = 'car' 
      AND target_vehicle_count >= 3; 
    \end{lstlisting}}\\ 
    \textbf{Q5} & 
    {\begin{lstlisting}[style=SQLStyle]
    SELECT frames FROM UA-DETRAC 
    WHERE vehicle_type = 'van'  
      AND target_vehicle_count >= 1; 
    \end{lstlisting}} \\ \bottomrule 
    \end{tabular}
    \caption{\textbf{Queries used for evaluating \sys}.}
    \label{tbl:queries}
\end{table}

\begin{table*}[]
    \begin{tabular}{cccccccc}
    \multicolumn{1}{c}{\textbf{Video Name}} & \textbf{Object Name} & \textbf{Object Count} & \textbf{Resolution} & \textbf{FPS} & \textbf{\# of Eval Frames} & \textbf{\# of TRUE Frames} & \multicolumn{1}{l}{\textbf{\% of TRUE Frames}}     \\ \hline
    UA-DETRAC                                & CAR                 & 1                      & 960x540             & 25           & 83791                        &            80744                &                    96.36\%                               \\
    UA-DETRAC                                & CAR                 & 2                      & 960x540             & 25           & 83791                        &            75249            &                        89.8\%                          \\
    UA-DETRAC                                & CAR                 & 3                      & 960x540             & 25           & 83791                        &               66485              &                       79.3\%          \\
    UA-DETRAC                                & VAN                 & 1                      & 960x540             & 25           & 83791                        &            25630                &                    30.5\%                            \\
    Seattle                                  & CAR                 & 1                      & 728x478             & 60           & 100k                       &              20731        &                            20.73\%                     \\
    Seattle                                  & CAR                 & 2                      & 728x478             & 60           & 100k                       &               1801        &                              1.8\%                       
    \end{tabular}
    
    \caption{\textbf{Datasets} -- 
    The key properties of the datasets used in our evaluation.
    We list the number of evaluated frames and the percentage of frames
    containing target object (\ie number of \cc{TRUE} frames).  
    }
    \label{tb:dataset}
\end{table*}

\PP{Datasets and Queries}   
We evaluate \sys on two datasets: (1) UA-DETRAC~\cite{wen2020uadetrac}, and (2)
Seattle~\cite{seattle}.
The key properties of these traffic-surveillance datasets are summarized
in~\cref{tb:dataset}.
UA-DETRAC consists of numerous short traffic camera videos (each of one minute
duration).
The latter dataset is a 30 minute long video of a traffic intersection in
Seattle.

We use the queries shown in \autoref{tbl:queries} to evaluate \sys and
other sampling algorithms.
We use SSD~\cite{liu2016ssd} as a reference inference model. 
We rank all the algorithms based on how well the labels that they assign to the
frames in the video compare against those assigned by SSD.
We use the label propagation technique shown in~\cref{fig:label-propgation}.

\PP{Sampling Algorithms} 
We compare six sampling algorithms in our evaluation:
(1) \none,
(2) \iframe, 
(3) \uniform,
(4) \noscope,
(5) \tasti,  
(6) \sysvgg, and
(7) \sys.
Here, \none refers to disabling the sampling optimization 
(\ie evaluating all the frames of the video).
It serves as an upper bound on accuracy (and a lower bound on latency).
\iframe refers to using the original I-frames picked by the x264 video
compression algorithm \cite{x264}.
Since the I-frame always comes at the beginning of the GOP, we propagate the
label assigned to this frame to the latter B- and P-frames within the same GOP.
\uniform refers to picking one frame out of every $k$ frames depending on
the number of required frames.

\noscope samples frames based on detecting difference between a predefined frame and current frame.
While two different methods are mentioned in the paper for determining the predefined frame, 
we use the method that detects the difference against an earlier frame of $t_{diff}$ seconds in the past.
~\cite{kang2017noscope}.
\tasti uses triplet loss on labeled video data to derive features from the video frames. 
The system then applies furthest point first (FPF) algorithm~\cite{fpf} and a
variant of K nearest neighbors~\cite{altman1992knn} for propagating the
labels to the remainder of the frames.
Since we focus on the unsupervised setting, we use the pretrained version of the
TASTI (referred to as TASTI-PT in~\cite{kang2020tasti}).

\sys utilizes all the optimizations outlined in this paper including:
(1) a custom feature extractor,
(2) temporal constraint,
(3) middle frame selection, and
(4) a custom encoder and decoder.
In contrast, \sysvgg uses all of these optimizations except that it utilizes a
pre-trained VGG-16 network.
Unless specified otherwise, all the sampling algorithms are configured to pick
the same number of frames for a fair comparison.

\PP{Hardware Environment} 
We perform the experiments on a server with these specifications:
\begin{itemize}
    \item CPU: 16 Intel(R) Xeon(R) Gold 6134 @ 3.20GHz
    \item GPU: 4 Geforce RTX 2080 Ti
    \item RAM: 385~GB
\end{itemize}
SSD runs at 30~fps on one RTX 2080 Ti GPU in this server. 

\subsection{Query Accuracy}
\label{ssec:query-accuracy}

\begin{figure*}
  \includegraphics[width=\textwidth]{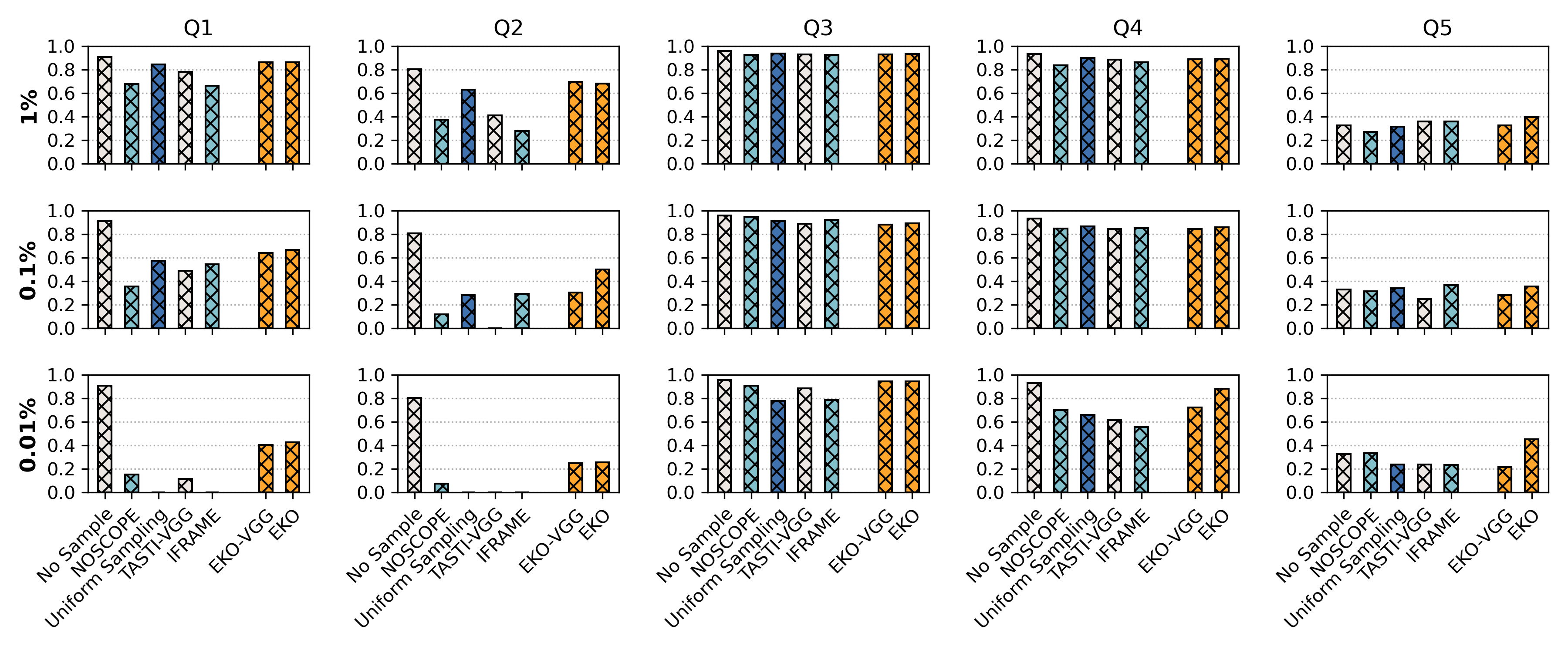}
  \caption{
  \textbf{Accuracy} -- 
  Comparison of the F1-score of different
  sampling algorithms across different queries.
  \sys strikes a balance between precision and recall metrics to achieve a high
  F1-score.
  }
  \label{fig:accuracy}
\end{figure*}

We first examine the end-to-end accuracy of \sys{'s} unsupervised, 
dynamic sampling algorithm outlined in~\autoref{sec:preprocess}.
We compare it against the other sampling algorithms listed
in~\autoref{sec:baselines}.  
In this experiment, \sys first loads all of the frames into memory and then 
selects a subset of frames using the given sampling algorithm.
It then sends those frames to the \execution that runs inference using
SSD~\cite{liu2016ssd}.
Lastly, the \execution propagates the output label for the sampled frame to 
other frames in the same cluster using the method illustrated
in~\cref{fig:label-propgation}.

\PP{Selectivity}
We vary the selectivity of the sampling algorithm (\ie the number of samples
picked) from 1\% to 0.01\%.
For example, with a dataset of 100~K frames and selectivity of 1\%, 
each algorithm is tailored to pick 1~K frames.
We configure all the algorithms to return the same number of samples to ensure 
a fair comparison.
We quantify query accuracy by measuring the F1-score.
It is important to strike a balance between the precision and recall metrics
for a given sampling algorithm to work well for diverse queries.
We compare the efficacy of the algorithm against \none.

We first configure the selectivity of all the algorithms to 1\%.
The results shown in~\cref{fig:accuracy}.
The most notable observation is that \sys{'s} F1-score only drops by 0.12
F1-points compared to \none. 
It is consistently better or on par with the second best sampling algorithm for
each query.

\begin{figure*}[t]
  \includegraphics[width=\textwidth]{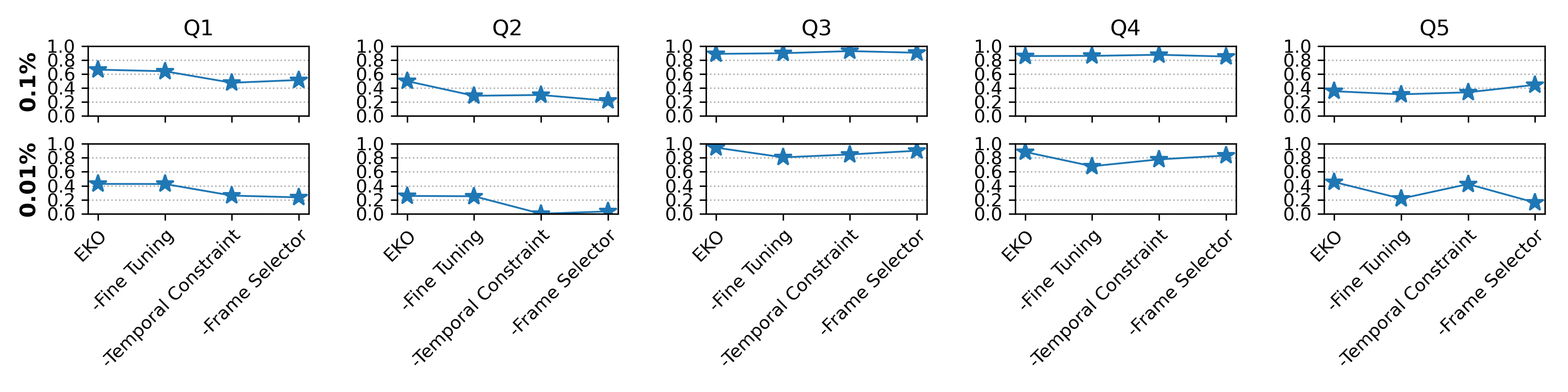}
  \caption{
  \textbf{Ablation Study} --
  Impact of different optimizations in \sys on accuracy.
  We disable optimizations one at a time:
  (1) feature extractor,
  (2) frame selector, ands
  (3) temporal constraint.
  }
  \label{fig:accuracycomponent}
\end{figure*}

On \scarone, \sys outperforms \uniform by 0.02 and \tasti by 0.18
points.
The reasons are twofold.
First, it picks a more representative set of frames that are not uniformly
spread across the video.
Second, it enforces a temporal constraint that improves the relevance of frames
mapped to the same cluster.

On \ucartwo, the difference between the sampling algorithms is minimal.
For instance, compared to \scarone, the gap between \sys and \uniform reduces 
from 0.02 to 0.0.
This is because the frequency of the target object in this dataset is high.
So, it is a relatively easier query to answer.
Even if the sampling algorithm picks a less representative frame, it is likely
that even this frame will contain the target object.
So, the impact of the sampling algorithm on F1-score is minimal.

On \uvanone, \sys outperforms the \none baseline by 0.07 points.
We attribute this to the comparatively low accuracy of SSD model on this query.
The query focuses on a van object that is harder to detect.
\sys picks a robust, representative subset of frames that reduces noise,
thereby enabling it to outperform \none.

\sys outperforms \sys-VGG on average by 10\% across all queries.
This is because the fine-tuned network allows \sys to better extract features.
The gap between \sys and \sys-VGG increases when we reduce the 
number of samples that they must pick.

\PP{Impact of Selectivity}
As shown in~\cref{fig:accuracy}, we observe that the drop in \sys{'s} accuracy
is minimal when we select fewer samples to process the query.
When the selectivity is reduced from 1\% to 0.01\%, \sys{'s} F1-score only 
drops by 9\%.
In contrast, accuracy of \uniform and \tasti drops by 53\% and 26\%,
respectively. 
We attribute this difference to the features generated by the network in \sys
and its dynamic sampling algorithm.

The accuracy gap between \sys and \none decreases as we increase the number of
samples.
On \scartwo, the gap shrinks from 68\% to 38\% when selectivity is increased
from 0.01\% to 1\%.
This is because \scartwo focuses on a rare event (\ie frames in the Seattle
dataset with more than two cars).
When the selectivity is reduced to 0.01\%, the difference between \sys and
\uniform is more prominent.

\PP{Complexity of Query}
The complexity of the query depends on:
(1) the frequency of occurence of the target object, and 
(2) the characteristics of the dataset.
For instance, while the object of interest in \uvanone (\ie van) appears more
frequently compared to that in \scarone (car), all of the sampling algorithms find it
hard to pick representative frames.
The reasons are twofold.
First, it is harder for the \udf to detect vans as opposed to cars.
This is why the F1-score of the \none algorithm is 64\% lower on \uvanone compared to
that in \scarone.
Second, the UA-DETRAC dataset's environment is more complex compared to 
that of the Seattle dataset.
The frames in the former dataset contain both vans and cars.
So, the sampling algorithm must carefully discriminate between the movement of
these objects.
Furthermore, this dataset consists of a collection short video clips focused on 
different traffic intersections.
So, it exhibits more significant variation in object movement compared to the
Seattle dataset that is a single long video focused on the same traffic
intersection.


\subsection{Ablation Study}
\label{ssec:ablation-study}

We next examine the impact of different optimizations in \sys on accuracy.
\sys uses three optimizations:
(1) a custom feature extractor,
(2) a custom frame selector, and
(3) a temporal constraint.
We disable these optimizations one at a time to quantify their impact on
accuracy.
In this experiment, we vary the selectivity of the algorithms from 0.1\% to
0.01\%.
Without the custom feature extractor,  \sys uses a pretrained VGG-16 network
(\ie same as \sys-VGG).
Without the temporal constraint, \sys uses a vanilla agglomerative ward
clustering algorithm to group the frames.
Without the custom frame selector, \sys picks the first frame in each cluster as
the representative frame.
The results are shown in~\cref{fig:accuracycomponent}.
The most notable observation is that while all the optimizations contribute to
the improvement in accuracy, the temporal constraint is the most important
optimization.

\PP{Feature Extraction}
The impact of disabling the custom feature extractor is more significant when
the algorithm must pick fewer frames.
For instance, on \uvanone disabling this optimization leads to an
accuracy drop of 25\% and 50\% when selectivity is 0.1\% and 0.01\%, respectively.
This illustrates the importance of finding more representative frames using a
better feature extractor.

\PP{Temporal Constraint}
The most important optimization in \sys is the temporal constraint.
The impact of this optimization is most prominent on complex queries 
(\ie \scarone, \scartwo).
The temporal constraint allows \sys to utilize the motion information inherently
present in the video.
So, the algorithm is robust enough to handle the rare events that these quries
seek to find.
It is less susceptible to confusing a black truck and a black car that are
present in completely different sections of the video.  
The impact of disabling these optimizations is minimal on \ucartwo,
\ucarthree, and \uvanone.
The frequencies of the target objects in these queries are so high that only
enabling the feature extractor and picking the middle frame is sufficient to
obtain quality results.
We further discuss the effect of temporal constraint in~\autoref{ssec:temporal-constraint}.

\PP{Frame Selection}
The impact of disabling the frame selector is most significant on \scartwo.
This query focuses on a rare event.
In this case, the middle frame is more likely to be representative of the
cluster as opposed to the first frame.
In constrast, on \ucartwo, \ucarthree, and \uvanone, 
disabling the frame selector often results in better accuracy. 
This is because the query focuses on a frequent event and an easier dataset.
So, the \execution is capable of correctly answering the query even using 
the first frame.
We further discuss the effect of frame selection strategy in~\autoref{ssec:frame-selection}

\subsection{Execution Time}
\label{ssec:query-speed}

\begin{figure}[t]
  \centering
  \includegraphics[width=0.4\textwidth]{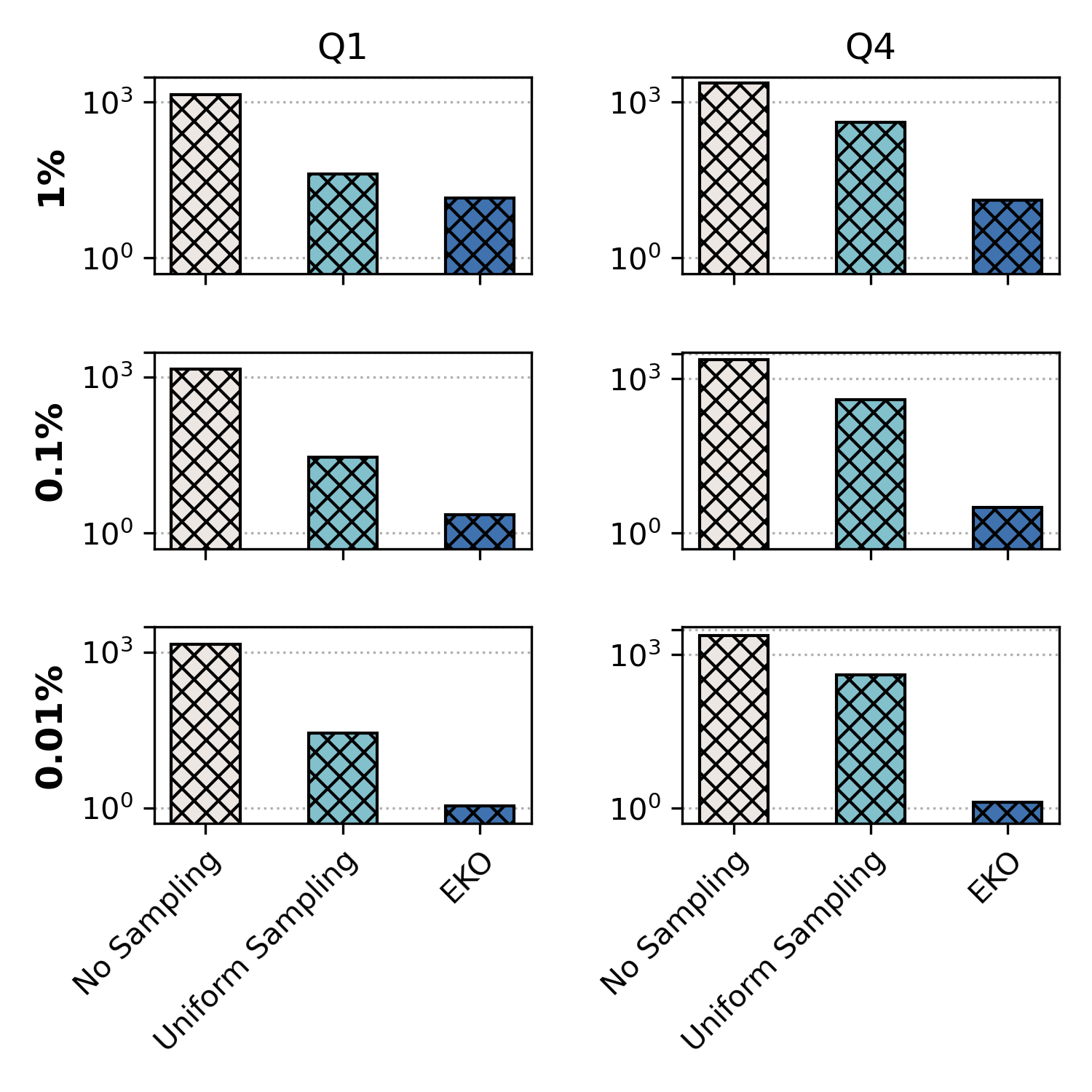}
  \caption{
  \textbf{Execution Time --}
  Comparison of the query processing time of different sampling algorithms.
  }
  \label{fig:speed}
\end{figure}

\begin{figure}[t]
  \centering
  \includegraphics[width=0.4\textwidth]{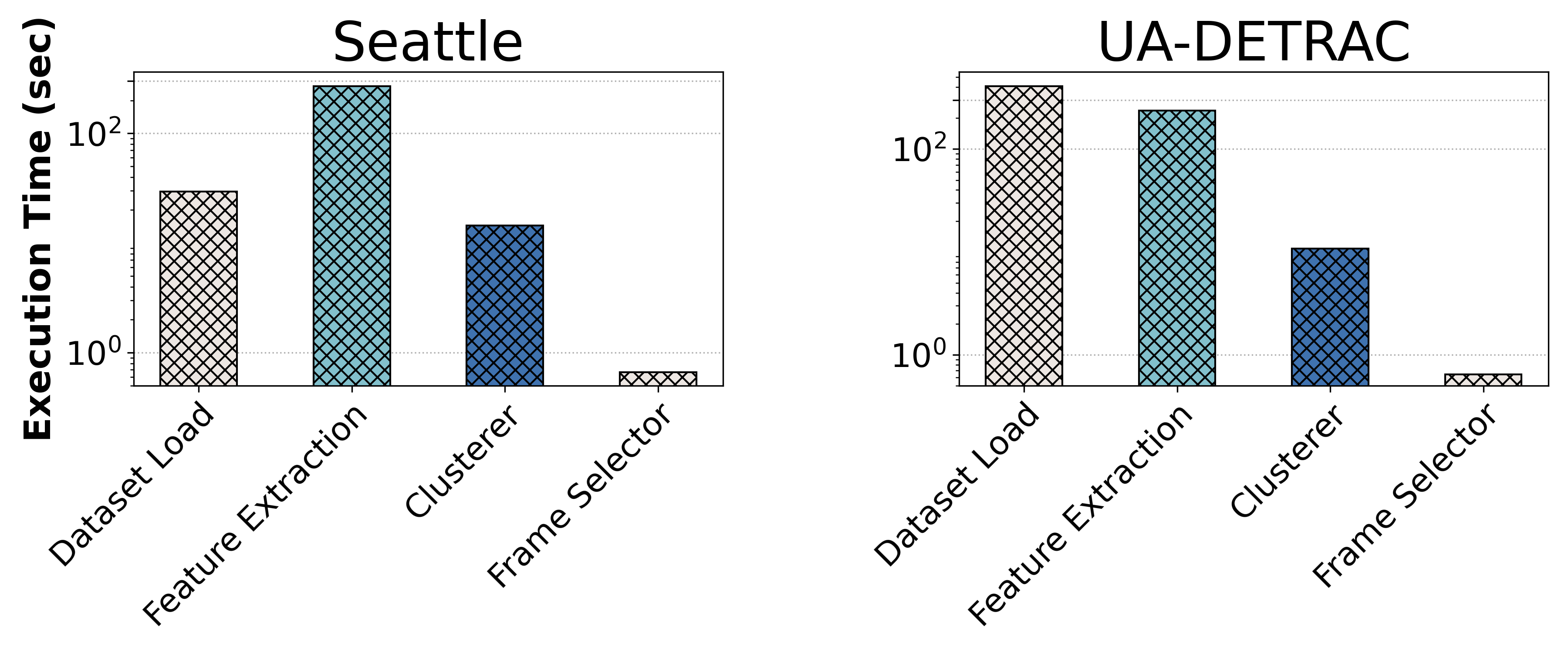}
  \caption{
  \textbf{Preprocessing Time --}
  Breakdown of time spent on different components of \preprocessor.
  }
  \label{fig:preprocessing-time}
\end{figure}

In this experiment, we compare the time taken by \sys to process queries in
comparison to \none and \uniform algorithms.
We focus on two representative queries: \scarone and \ucartwo.
With \sys, we assume that the \preprocessor has already constructed a compressed
video with meta-data about the representative frames.
Since this is an one-time cost, we do not include the pre-processing time in the
results.
We measure the time taken to load the sampled frames from disk to memory, and to
perfor inference on those frames.
We vary the selectivity of the algorithms 1\% to 0.01\%.

The results are shown in \cref{fig:speed}.
On \scarone, for selectivity 1\%, \sys executes the query 3$\times$ and 100$\times$ compared to \uniform
and \none, respectively.
Comparative speed increases as we decrease the numbers of samples used to evalute the query.
We attribute this speedup to two factors.
First, \sys combines the adaptive sampling algorithm and the compression scheme
outlined in~\autoref{sec:encoding-decoding}.
With both \uniform and \none, all of the frames in the video are loaded from
disk to memory.
In contrast, the custom decoder in \sys allows it to only load the desired
number of sampled frames from the video into memory.
Second, unlike \none, \sys and \uniform only run inference on the sampled frames
and propagate the labels to the other frames.
On average, it takes 0.3 ms and 2.7 ms to load a frame from disk to
memory and to run inference on a given frame, respectively.

On \ucartwo, \sys for selectivity 1\% executes the query 31$\times$ and 180$\times$ compared to \uniform
and \none, respectively.
The impact of \sys is more prominent on this query focusing on the UA-DETRAC
dataset.
This is because this dataset is organized as a a series of images (unlike the
Seattle dataset which is a compressed video format).

\PP{Impact of Selectivity}
We next reduce the selectivity of the algorithms from 1\% to 0.01\%.
On \scarone, \sys executes the query 27$\times$ faster than \uniform.
The speedup is 9$\times$ higher than that observed when the selectivity is set
to 1\%.
This experiment illustrates the importance of picking a subset of important
frames that are sufficient to answer the query with high confidence.

\PP{Pre-Processing Time}
We next report the time taken to pre-process the video.
This is an one-time cost and is amortized across multiple queries on the same
dataset.
The results are shown in~\cref{fig:preprocessing-time}.
We split the preprocessing time into four components: 
(1) dataset loading,
(2) feature extraction, 
(3) clustering, and 
(4) frame selection.
With the Seattle dataset, feature extraction was the most time consuming
component (270~s).
%
%
With the UA-DETRAC dataset, bottleneck lies in the data loading step.
This is because the dataset is organized a series of frames.
%

\subsection{Memory and Storage Footprint}
\label{ssec:storage}

\begin{figure}[t]
\includegraphics[width=0.49\textwidth]{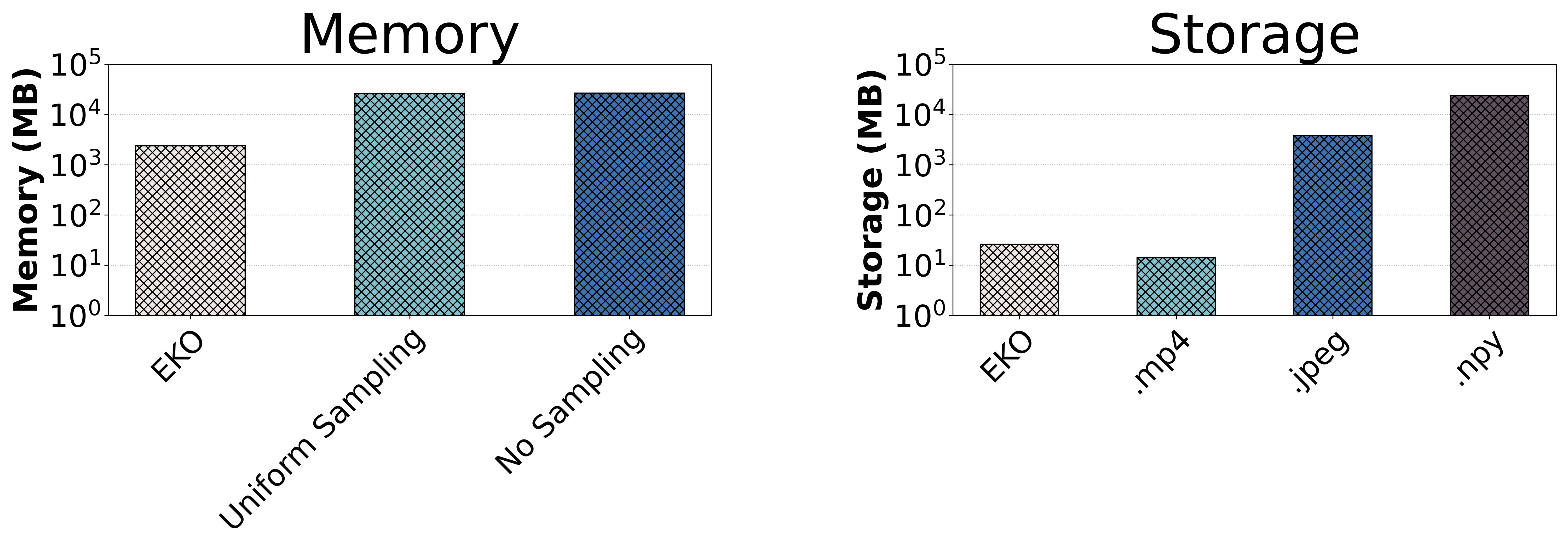}
  \caption{
  \textbf{Memory and Storage Footprint --} 
           (a) Memory footprint of decompressed video, and
           (b) Storage footprint of compressed video on disk.
         }
  \label{fig:eko-memory-experiment}
\end{figure}

In this experiment, we quantify the memory and storage footprint of \sys.
We focus on \scarone and configure the selectivity to 1\%.
We compute storage footprint based on the size of the Seattle dataset.
The results are shown in~\cref{fig:eko-memory-experiment}.
We refer to the re-encoded video as \sys.
\textsc{mp4} refers to the video emitted by the $h.264$ codec.
\textsc{jpeg} refers to storing the entire video as series of JPEG images.
\textsc{npy} refers to the numpy representation for storing raw video content.

\PP{Impact on Memory}
\sys consumes 2.37~GB of CPU memory while other algorithms consume 26~GB.
This is because with the latter algorithms, the VDBMS decodes the entire video
in memory.
So, \sys is $~10\times$ memory efficient compared to \uniform and \none.
The raw size of this dataset is only 240~MB.
The rest of memory footprint stems from loading the SSD model.
This experiment illustrates that \sys scales better to large datasets.

\PP{Impact on Storage}
\sys's re-encoded video takes 80\% more space compared to \textsc{mp4}.
The reasons are twofold.
First, \sys contains more I-frames compared to the \textsc{mp4} representation.
Second, it generates metadata to faciliate dynamic sampling.
\sys is $100\times$ and $270\times$ more space efficient compared to
widely-used \textsc{jpeg} and \textsc{npy} formats, respectively.

\subsection{Impact of Temporal Constraint}
\label{ssec:temporal-constraint}

\begin{figure*}
  \includegraphics[width=0.95\textwidth]{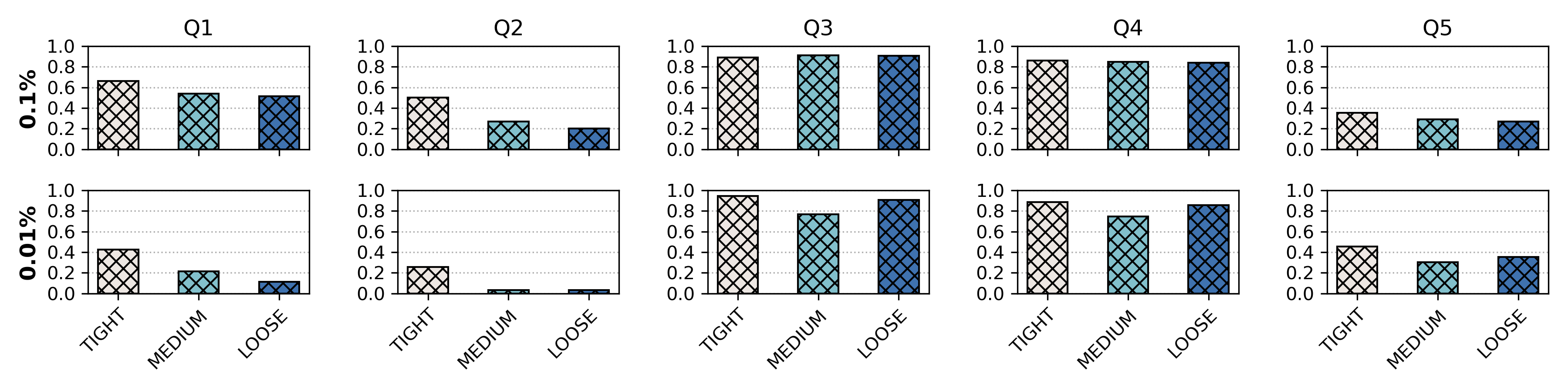}
  \caption{
  \textbf{Impact of Temporal Constraint} --
  Comparison of accuracy across different types of temporal constraints:  
  (1) \tight, (2) \medium, and (3) \loose.
  }
  \label{fig:eko-temporal-constraint}
\end{figure*}

\begin{figure*}
  \includegraphics[width=0.95\textwidth]{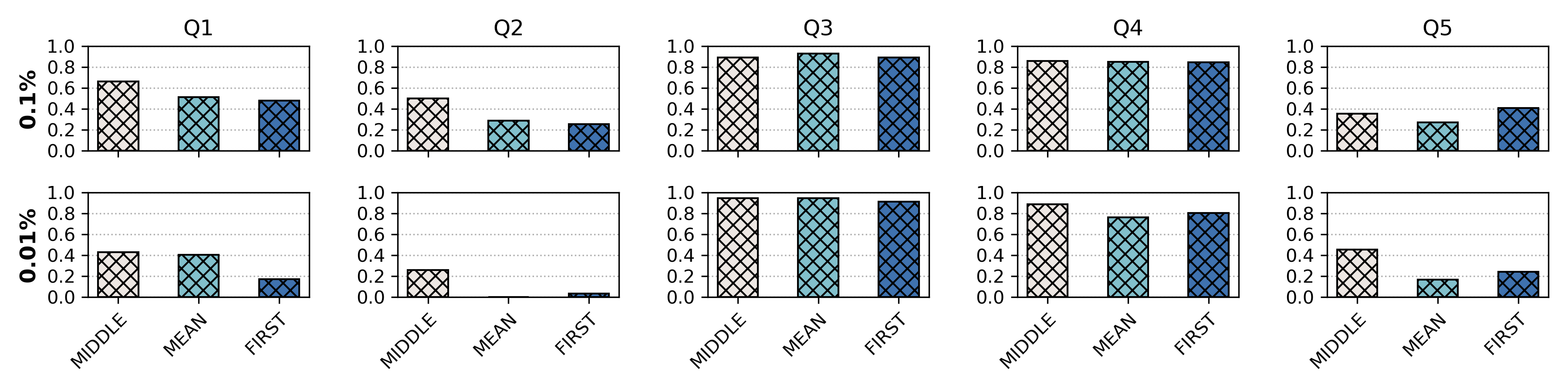}
  \caption{
  \textbf{Impact of Frame Selection} --
  Comparison of accuracy across different types of frame selection policies.}
  \label{fig:selection-method}
\end{figure*}

We next examine how the temporal constraint parameter affects the query accuracy
of \sys for selectivities of 0.1\% and 0.01\%.
We consider three configurations.

\squishitemize
\item  \tight denotes a strong temporal constraint. 
It only allows the cluster to be formed by connecting temporally adjacent
frames in the video.
It is the default configuration in other experiments.

\item \loose represents a much weaker temporal constraint.
Clusters may be formed within a temporal span of 100 frames.

\item \medium represents a configuration in between these two extremes.
The clusterer may be form clusters with a temporal span ranging up to 50
frames.

\squishend

The results are shown in~\cref{fig:eko-temporal-constraint}.
The most notable observation is that relaxing the temporal constraint leads to a
significant accuracy drop in many queries.
With \scarone (selectivity = 0.1\%), \tight outperforms \loose and \medium
by 14\% and 19\%.
With \scartwo, the gap increases to 48\% and 65\%, respectively
This shows that the algorithm performs better under a tight temporal constraint.
On \ucartwo and \ucarthree, \sys works well under all the three constraints.
This is because this query focuses on a frequent event.
Since most frames satisfy the query's constraint, the impact of the feature
extractor in \sys is reduced.
The importance of the temporal constraint increases when we reduce selectivity
to 0.01\%.
When \sys must pick a smaller set of samples, it must effectively utilize the
motion information present in the video.

\subsection{Impact of Frame Selection}
\label{ssec:frame-selection}

We next examine the impact of frame selection on accuracy.
In this experiment, we configure \sys to use the custom feature extractor and
place a \tight temporal constraint to isolate the effect of frame selection.
In particular, we examine three policies for selecting frames within \sys:
(1) \first,
(2) \mean, and
(3) \midd.
With \first, \sys picks the first frame within each cluster while sampling
frames.
This is based on how I-frames are selected in canonical storage
formats~\cite{avi1998}.
With \mean, \sys computes the mean of all the frames within the cluster and
picks the frame that is closest to the cluster's mean frame.
Lastly, \midd refers to picking the middle frame within each cluster from a
temporal standpoint.
We vary the selectivity from 0.1\% to 0.01\%.
The results are shown in~\cref{fig:selection-method}.

\PP{\midd vs \first}
On average, \midd outperforms \first by 27\% across all queries.
The gap between \midd and \first is more prominent on queries focusing on rare
events.
Given the continous motion in videos, the middle frame  is more representative
of other frames in the cluster as opposed to the first frame.
For instance, assume that the size of a given cluster is $n$ frames.
The \first policy leads to a maxmimal difference of $n$ frames (\ie distance
between the first and the last frames in the cluster).
In contrast, the \midd policy leads to a maxmimal difference of $\frac{n}{2}$
frames.
This variation in the frame selection policy is the key difference between
\uniform and \iframe in~\cref{fig:accuracy}.

\PP{\midd vs \mean}
On average, \midd outperforms \mean by 26\% across all queries.
This is because the \mean policy does not work well with the continous motion in
videos.
Consider a cluster of frames wherein a car is moving from bottom left of the
frame to the top right.
The mean of this cluster would be a blurry smear of a car moving across the
frame.
While returning the frame closest to this mean frame, the \mean policy is likely
to pick a non-representative frame.
This experiment illustrates the importance of using the \midd policy in
\sys.

\section{Related Work}
\label{sec:related-works}

We now outline prior work in the areas of: 
(1) video analytics, 
(2) unsupervised image feature extraction, and 
(3) video storage and compression techniques.

\PP{Video Analytics}
Researchers have presented several techniques for accelerating binary
classification queries over videos.
They are based on model cascades~\cite{kang2017noscope,anderson2019tahoma},
lightweight filters~\cite{lu2018pp}, and specialized
models~\cite{hsieh2018focus}.
\noscope~\cite{kang2017noscope} uses a difference detector for filtering out
irrelevant frames and uses a specialized model for faster inference.
\tahoma~\cite{anderson2019tahoma} utilizes a cascade of fast, high precision
image classifiers.
The lightweight models earlier in the cascade are able to quickly answer
easier queries.
\probabilistic~\cite{lu2018pp} are lightweight models that filter out images
that are not likely to satisfy the query.
Instead of using high precision models as done by \tahoma, it uses models with
high recall.
\focus uses specialized models to pre-process videos.
By grouping frames with similar objects and using expensive models only for
harder-to-process frames, it reduces query execution time.

Another line of research focuses on optimizing specific types of queries.
\blazeit~\cite{kang2018blazeit} uses control variates to optimize aggregate and
limit queries by reducing the number of samples required to meet the error 
bound supplied by the user.
\miris~\cite{bastani2020miris} speeds up object tracking queries by varying the
sampling rate based on the contents of the video.
\exsample~\cite{moll2020exsample} illustrates the importance of picking more
samples in portions of the video that are more likely to contain objects of
interest.
%


\PP{Unsupervised Image Feature Extraction}
There are several unsupervised algorithms for extracting features from images
using~\cite{dec,decjoint,deepcluster}.
DEC uses clustering for this task~\cite{dec}.
It updates its extraction network by jointly updating the cluster centers 
and the DNN parameters while minimizing the KL divergence between the 
data points and the desired target distribution.
DEPICT~\cite{decjoint} improves upon DEC by introducing regularization terms in
the loss function for better clustering results.
It also updates the network's parameters to minimize the KL divergence between
the data points and the target variable.
DeepCluster~\cite{deepcluster} differs from these algorithms in that it utilizes
a multi-nomial logistic loss function along with regularization.

\sys differs from prior algorithms in that it does not compute loss based on 
the generated cluster labels.
Instead, it seeks to minimize the difference between the generated features and
desired features.
As discussed in~\autoref{sec:preprocess}, this optimization is critical for
videos wherein a given image may be assigned different labels based on the
applied temporal constraint.
Using the labels to compute the loss function would reduce the efficacy of the
network.

\PP{Video Storage and Compression}
Researchers have presented several systems for managing
videos~\cite{haynes2018lightdb,daum2020tasm,xu2019vstore}.
\vstore periodically updates the parameters that determine how the video is
stored (\eg image quality, crop factor, resolution, sampling rate) based on 
the feedback from the consumers of the video.
While \vstore modifies the sampling rate across videos, it does not dynamically
vary it within the video.  
\tasm supports fast spatial, random access using a tile-based compression
scheme.
\lightdb focuses on managing virtual, augmented, and mixed reality videos.
These videos consists of light fields that are higher-dimensional than 
traditional videos.
\sys differs from these systems in that it seeks to leverage the temporal
redundancy in the videos.
Prior efforts have applied deep  learning for the video compression
problem~\cite{lu2019dvc,rippel2019learnedvc,wu2018video}.
Specifically, one line of research focuses on minimizing distortion by
maintaining as much motion information as possible compared to 
traditional video compression techniques~\cite{lu2019dvc,wu2018video}.
Another line of research is centered on compression networks that optimize 
for storage space while maintaining the same level of distortion compared to 
traditional compression techniques~\cite{rippel2019learnedvc}.
\sys differs from these efforts in that it is tailored for answering video
queries.
It generates more accurate results by combining adaptive sampling with the
compression scheme.

\section{Limitations and Future Work}
\label{sec:future-work}

\sys currently only supports binary classification queries similar to many
other VDBMSs~\cite{kang2017noscope,anderson2019tahoma,lu2018pp,hsieh2018focus}.
It does not support more complex queries like object localization.
\begin{lstlisting}[style=SQLStyle]
    SELECT target_object_bounding_boxes 
    FROM UA_DETRAC 
    WHERE vehicle_type = 'car' 
      AND target_vehicle_count >= 2;
\end{lstlisting}

We could leverage the clustering and label propagation algorithms in \sys to
tackle this problem.
After clustering the frames with temporal constraints, we could extend \sys to
derive the movement vectors within each generated cluster during the offline,
video ingestion phase.
Then, during online query processing, \sys will leverage this meta-data to
propagate the bounding boxes to the other frames within the cluster.
%

\section{Conclusion}
{\label{sec:conclusion}}

In this paper, we presented \sys, a storage engine for efficiently managing
video data.
\sys leverages an unsupervised sampling algorithm for picking important frames
in videos.
Its custom feature extractor enables it to outperform the state-of-the-art
sampling algorithms with respect to accuracy on diverse queries and datasets.
\sys contains a novel encoder and decoder for efficiently retrieving the key
frames while processing queries.
This allows the \execution to fetch only the relevant frames without decoding
the entire video.
We show that \sys reduces query execution time by 3$\times$ and memory
footprint by 10$\times$ in comparison to state-of-the-art VDBMSs.

\clearpage
\newpage
{
\bibliographystyle{ACM-Reference-Format}
\small
\raggedright
\balance
\bibliography{eko}
}


\end{document}
\endinput